\documentclass[a4paper,11pt]{article}
\pdfoutput=1 

\usepackage{jheppub} 

\usepackage[T1]{fontenc} 
\usepackage{multirow}
\usepackage{caption}

\newcommand{\hp}{\mathbf{\hat p}}
\newcommand{\hk}{\mathbf{\hat k}}

\newcommand{\hr}{\mathbf{\hat r}}
\newcommand{\hn}{\mathbf{\hat n}}
\newcommand{\ha}{\mathbf{\hat a}}
\newcommand{\hb}{\mathbf{\hat b}}

\newcommand{\ttbar}{t{\bar t}}
\newcommand{\mttbar}{M_{t{\bar t}}}

\newcommand{\Sp}{{\mathbf S}_t}
\newcommand{\Sm}{{\mathbf S_{\bar t}}}
\newcommand{\hlp}{{\mathbf{\hat\ell}_+}}
\newcommand{\hlm}{{\mathbf{\hat\ell}_-}}

\newcommand{\Hmu}{{\hat\mu}_t}
\newcommand{\Hd}{{\hat d}_t}

\preprint{TTK-15-16}

\title{\boldmath A set of top quark spin correlation and polarization observables for the LHC: Standard Model predictions and new physics contributions}

\author[a,1]{Werner Bernreuther,\note{Corresponding author.}}
\author[a]{Dennis Heisler}
\author[b]{and Zong-Guo Si}

\affiliation[a]{Institut f\"ur Theoretische Teilchenphysik und Kosmologie, RWTH Aachen University,\\52056 Aachen, Germany}
\affiliation[b]{School of Physics, Shandong University, Jinan, Shandong 250100, China}

\emailAdd{breuther@physik.rwth-aachen.de}
\emailAdd{heisler@physik.rwth-aachen.de}
\emailAdd{zgsi@sdu.edu.cn}

\abstract{We consider top-antitop quark $(\ttbar)$ production at the Large Hadron Collider (LHC) with
 subsequent decay into dileptonic and lepton plus jets final states. We present a set of leptonic angular correlations and distributions
 with which experiments can probe all the independent coefficient functions 
  of the top-spin dependent parts of the $\ttbar$ production spin density matrices. 
 We compute these angular  correlations and distributions for LHC center-of-mass energies  8, 13, and 14 TeV within the Standard Model
 at next-to-leading order in the QCD coupling including the mixed QCD-weak corrections and for 
  the transverse top-quark polarization and the $\ttbar$ charge asymmetry also the mixed QCD-QED corrections. 
  In addition we analyze and compute the effects of  new interactions on these observables
 in terms of a gauge-invariant  effective Lagrangian that contains all operators  relevant for
 hadronic $\ttbar$ production up to mass dimension six. }

\keywords{Hadron collider physics, top quark, spin, QCD, electroweak interactions, new physics, anomalous couplings,
CP violation. \\ PACS number(s): 12.38.Bx, 13.88.+e, 14.65.Ha}

\begin{document} 
\maketitle
\flushbottom

\section{Introduction}
\label{sec:intro}

All available experimental data on top quarks show that this particle behaves as a bare quark that does not hadronize.
 This fact implies that observables associated with or induced by top-quark spin effects can be used as reliable tools
  for investigating the interactions of this quark. As far as hadronic top-antitop quark $(\ttbar)$ production is concerned,
   the correlation of the $t$ and $\bar t$ spins and polarization of the $t$ and $\bar t$ samples in $\ttbar$ events were first experimentally
    explored at the 
    Tevatron \cite{Aaltonen:2010nz,Abazov:2011ka,Abazov:2011gi,Abazov:2012oxa}. 
    Evidence for  $\ttbar$ spin correlations was first reported by the D$\emptyset$ experiment \cite{Abazov:2011gi} and they were first observed 
     by the ATLAS experiment  \cite{ATLAS:2012ao}.
    The ATLAS and CMS experiments at the LHC (7 and 8 TeV) have performed a number of top spin correlation and polarization 
     measurements, using various sets of observables
    \cite{ATLAS:2012ao,Aad:2013ksa,Chatrchyan:2013wua,Aad:2014pwa,CMS:2014bea,Aad:2014mfk,CMS:2015dva}.
    These measurements have reached a level of precision that allow for a meaningful interpretation. 
    All measurements of $\ttbar$ spin correlations and top-quark polarization so far are in agreement with respective Standard Model (SM) 
     predictions \cite{Bernreuther:2001rq,Bernreuther:2004jv,Bernreuther:2010ny,Bernreuther:2013aga}.
  
   The degree of top-quark polarization  in a $\ttbar$ sample and the strength of the correlation of the $t$ and $\bar t$ spins
     depends on the production dynamics and, for a 'given' production dynamics, on the choice of observable, that amounts to a choice of
      reference axes. (These axes may be interpreted as $t$ and $\bar t$ spin quantization axes.) In order that future measurements at the LHC will
      be able to extract complete information 
       about the $\ttbar$ production dynamics from their data as far as top-spin effects are concerned, it is desirable to devise
       a set of observables, in some basis,  which probe all of the independent coefficient functions of the top-spin dependent parts 
        of the $\ttbar$ production spin density matrices.
        In this paper we present such a set. The members of this set are constructed such that they have definite properties
         with respect to discrete symmetries. For definiteness we consider, as 
          far as  $\ttbar$ spin correlation effects are concerned, the   dileptonic $\ttbar$ decay modes and,
           as far as polarization effects are concerned, also the lepton plus jets final states. These modes  are the most sensitive
          channels for the investigation of top-spin effects. We compute the distributions and expectation values of these observables 
          at next-to-leading order (NLO) QCD with respect to $\ttbar$ production and decay and take into account also the NLO mixed QCD weak-interaction contributions
          and for the transverse $t$ and $\bar t$ polarizations also the mixed QCD-QED corrections of
           order $\alpha_s^2\alpha$.  In addition,  we assess the sensitivity of our set of spin correlation and polarization observables to new physics (NP)
            effects in the framework of  the effective field theory approach. We take into account all gauge invariant operators of dimension six which are relevant for 
             hadronic $\ttbar$ production \cite{Buchmuller:1985jz,Grzadkowski:2010es,AguilarSaavedra:2008zc,AguilarSaavedra:2009mx,Zhang:2010dr,AguilarSaavedra:2010zi,Degrande:2010kt}. 
             We compute the contributions of the associated anomalous couplings to our observables. 
              There is already an extensive literature on phenomenological studies, including
       \cite{Atwood:1992vj,Brandenburg:1992be,Atwood:1994vm,Haberl:1995ek,Cheung:1995nt,Cheung:1996kc,Grzadkowski:1997yi,Yang:1997iv,Hikasa:1998wx,Zhou:1998wz,Atwood:2000tu,Martinez:2001qs,Lillie:2007hd,Gupta:2009eq,Gupta:2009wu,Hioki:2009hm,HIOKI:2011xx,Choudhury:2009wd,Cao:2011hr,Kamenik:2011dk,Bach:2012fb,Gabrielli:2012pk,Hioki:2012vn,Englert:2012by,Biswal:2012dr,Baumgart:2012ay}
   and \cite{Bernreuther:2013aga,Ayazi:2013cba,Hioki:2013hva,Fabbrichesi:2013bca,Belanger:2012tm,Hayreter:2013kba,Konig:2014iqa,Aguilar-Saavedra:2014iga,Choudhury:2014lna,Kiers:2014uqa,Prasath:2014mfa,Tweedie:2014yda,Carmona:2014gra,Zhang:2014rja,Hesari:2014hva,Kobakhidze:2014gqa,Hioki:2014eca,Jung:2014kxa,Gaitan:2015aia,Godbole:2015bda,Franzosi:2015osa,Buckley:2015nca,deBlas:2015aea},
                     on how anomalous couplings, in particular, a  non-zero
               chromo-magnetic and/or chromo-electric dipole moment of the top quark would affect $\ttbar$ production at hadron colliders.
               In fact, if one considers only the two chromo-moments and puts all other anomalous couplings to zero, it suffices to take only 
                the experimental results for  the $\ttbar$ cross section at the Tevatron 
                and the LHC (7 and 8 TeV)  and the respective NNLO QCD predictions \cite{Czakon:2013goa} in order to obtain  sensitive bounds on 
                these two anomalous couplings \cite{Hioki:2013hva,Aguilar-Saavedra:2014iga,Franzosi:2015osa}.
                 The point we want to make with this part of our analysis is that our set of observables, including  
             the  $\ttbar$ charge asymmetry\footnote{The $\ttbar$ cross section should, of course, also
              be used but is, in principle, redundant; see section~\ref{sec:obsres}.} is  i) large enough to get information on 
            all  anomalous couplings of our effective Lagrangian (in fact the number of observables in this set exceeds the number of couplings) and ii) these
             observables allow to disentangle the various anomalous contributions to a considerable extent.
             
       This paper is organized as follows. In the next section we decompose the $\ttbar$ production density matrices of the reactions $gg, q{\bar q}\to \ttbar$
        in terms of an orthonormal basis and classify the coefficient functions of these matrices with respect to parity P, CP, naive time reversal $T_N$, and Bose
         symmetry. The decomposition applies to all parton reactions $ij\to \ttbar X$. 
          In section~\ref{sec:Lagreff} the effective Lagrangian ${\cal L}_{\rm NP}$ is introduced that we use for describing non-resonant new physics in 
          hadronic $\ttbar$ production. We use the operator basis employed in \cite{Degrande:2010kt}, albeit with different normalization conventions
           for the anomalous couplings. Possible new physics contributions to the $t\to W b$ decay 
            vertex do not affect our leptonic angular correlations and distributions defined in section~\ref{sec:obsres}.
             We briefly  recall the presently available bounds on the 
              anomalous couplings contained in ${\cal L}_{\rm NP}$ and discuss the domain of applicability of the linear approximation used in 
               section~\ref{sec:obsres}. In this section, we consider $\ttbar$ production and decay into dileptonic and lepton plus jets final
                states at the LHC. For the dileptonic final states, we  present a set of 
                leptonic angular correlation variables that are constructed such that each member of this set probes a different $\ttbar$ spin correlation
                 coefficient of the production density matrices. In addition, we list a set of single lepton distributions/observables which probe the 
                  $t$ and $\bar t$ polarization coefficients of the  production density matrices. These observables can be applied
                  both to  dileptonic and lepton plus jets final states.
                 We compute, for $\ttbar$ production at the LHC at 8, 13, and 14 TeV,  the two- and one-dimensional distributions associated with these
                  observables  at NLO QCD  ($\ttbar$ production and decay) including the NLO mixed QCD weak-interaction corrections, 
                   for the transverse $t$ and $\bar t$ polarizations  also the mixed QCD-QED corrections, 
                   and, in addition, the contributions induced by  ${\cal L}_{\rm NP}$ which are linear in the anomalous couplings. For completeness we
                    determine also the contributions of ${\cal L}_{\rm NP}$ to the $\ttbar$ cross section $\sigma_{\ttbar}$ and the $\ttbar$ charge asymmetry
                    at the LHC, $A_C$. For the observables considered in this paper we estimate the application range of the linear approximation
                     in the anomalous couplings.
                      Our results show that our set of correlation and polarisation observables, including the $\ttbar$ charge asymmetry $A_C$,
                     allows to determine all  anomalous couplings of ${\cal L}_{\rm NP}$ and disentangles these contributions to a considerable extent.
                   We conclude in section \ref{sec:sumconc}.

\section{$\ttbar$ production density matrices in an orthonormal basis}
\label{sec:ortho}
 We consider in this section, for the purpose of setting up our notation,
   hadronic $\ttbar$ production by  the  $2\to 2$ reactions
\begin{equation} \label{ggproc}
 g(p_1) + {g}(p_2) \to t(k_1, s_1) + {\bar t}(k_2, s_2) \, ,
 \end{equation}
\begin{equation} \label{qqproc}
 q(p_1) + {\bar q}(p_2) \to t(k_1, s_1) + {\bar t}(k_2, s_2) \, , 
\end{equation}
where $p_j, k_j$ and $s_1, s_2$ refer to the 4-momenta of the partons and to the
  spin 4-vectors of the $t$ and $\bar t$ quarks, respectively.
The production density matrices
 of these reactions are defined by
\begin{eqnarray}  R^{I}_{\alpha_1\alpha_2,\ \beta_1\beta_2}
 =  {\overline\sum} & \langle
t(k_1,\alpha_2), \bar t(k_2,\beta_2)  \vert {\cal T}\vert
 a(p_1),b(p_2) \rangle^* \nonumber \\
 & \times \langle t(k_1,\alpha_1), \bar t(k_2,\beta_1) \vert {\cal T}\vert
a(p_1),b(p_2) \rangle
\label{ggtt} 
\end{eqnarray}
 where $I\equiv a b = gg, q{\bar q}$ and the bar denotes averaging over
  the spins and colors of $I$ and summing over the colors of $t, \bar t$. 
   Moreover, $\alpha,\ \beta$ are spin labels referring to $t$ and $\bar t$, respectively.
The matrices $R^{I}$ can be decomposed
in the spin spaces of $t$  and $\bar{t}$ as follows:
 \begin{eqnarray} R^{I}= & f_I \left[ A^I 1\!{\rm l}\otimes 1\!{\rm l}+{\tilde B}^{I+}_{i}\sigma^i
\otimes 1\!{\rm l}+{\tilde B}^{I-}_{i}1\!{\rm l}
\otimes\sigma^i+{\tilde C}^I_{ij}\sigma^i\otimes\sigma^j \right] \, , \label{Rot} \\
  f_{gg} = & \displaystyle{\frac{(4\pi\alpha_s)^2}{N_c (N_c^2-1)} \, , \qquad f_{q\bar q}=\frac{(N_c^2-1)(4\pi\alpha_s)^2}{N_c^2} } \, , \nonumber 
\end{eqnarray}
 where $N_c$ denotes the number of colors. The first (second) factor in the tensor products of the $2\times 2$
unit matrix $1\!{\rm l}$ and of the Pauli matrices $\sigma^i$ refers to the $t$
 $(\bar{t})$ spin space.

The functions ${\tilde B}^{I\pm}_{i}$ and ${\tilde C}^I_{ij}$ can
be further decomposed, using an orthonormal basis which we choose as follows.  Here and in the following
$\hk$ denotes the top-quark direction of flight in the $\ttbar$ zero-momentum frame (ZMF).
 Here $\hp={\bf\hat p}_1$ denotes the direction of one of the incoming partons in this frame.
  Then the following set forms a right-handed orthonormal
   basis:
\begin{align} \label{orhtoset}
	\bigl\{\hr, \, \hk, \, \hn \bigr\}: \qquad \hr = \frac{1}{r}\bigl(\hp - y\hk\bigr),\quad
	\hn = \frac{1}{r}\bigl(\hp \times \hk \bigr),&\qquad
	y = \hk \cdot \hp, \; r = \sqrt{1-y^2}. 
\end{align}
 Using rotational invariance we decompose
  the 3-vectors ${\bf\tilde B}^{I\pm}$ and the
 $3\times 3$ matrices ${\tilde C}^I_{ij}$ (which have a symmetric and antisymmetric part with 6 and
 3 entries, respectively) with respect\footnote{The decomposition of the
  ${\tilde C}^I_{ij}$ is not unique because
 $\delta_{ij} = \hat n_i \hat n_j + \hat r_i \hat r_j + \hat k_i \hat k_j.$} to the  basis \eqref{orhtoset}:
\begin{align}
	{\tilde B}^{I \pm}_i &=    b^{I\pm}_r\, \hat r_i \;+\; b^{I\pm}_k\, \hat k_i  \;+\; b^{I\pm}_n \, \hat n_i \, , \label{eq:newB} \\
	{\tilde C}^I_{ij} &=\;  c^I_{rr}\,\hat r_i \hat r_j \;+\; c^I_{kk}\,\hat k_i \hat k_j \;+\; c^I_{nn}\,\hat n_i \hat n_j \notag \\
                    & +\; c^I_{rk}\,\bigl(\hat r_i \hat k_j \;+\; \hat k_i \hat r_j \bigr) 
                +\; c^I_{kn}\,\bigl(\hat k_i \hat n_j \;+\; \hat n_i \hat k_j \bigr) 
              \;  +\; c^I_{rn}\,\bigl(\hat r_i \hat n_j \;+\; \hat n_i \hat r_j \bigr) \notag\\
	&+\; \epsilon_{ijl}\,\bigl( c^{I}_r \, \hat r_l \;+\; c^{I}_k \, \hat k_l \;+\; c^{I}_n  \hat n_l   \bigr) \, . \label{eq:newC}
\end{align}
 The coefficients  $b^{I\pm}_v$,  $c^I_{vv^{\prime}}$  are functions of  the partonic c.m. energy  squared, $\hat s$,
  and of $y=\hk \cdot \hp$   which is equal to the  top-quark scattering angle $\cos\theta_t^*$.
 The terms in the antisymmetric part of \eqref{eq:newC} may also be represented as follows:
  \begin{equation} \label{antiCnew}
  c^{I}_r\epsilon_{ijl}\hat r_l =c^{I}_r(\hat k_i \hat n_j -\hat n_i \hat k_j), \quad 
  c^{I}_k\epsilon_{ijl}\hat k_l =c^{I}_k(\hat n_i \hat r_j -\hat r_i \hat n_j), \quad  
  c^{I}_n\epsilon_{ijl}\hat n_l =c^{I}_n(\hat r_i \hat k_j -\hat k_i \hat r_j).  
  \end{equation}
  Because the initial $gg$ state is Bose symmetric, the matrix $R^{gg}$ must satisfy
 \begin{equation}
  R^{gg}(-{\bf p}, {\bf k}) = R^{gg}({\bf p}, {\bf k}) \, .
  \label{bose}
  \end{equation}
  If CP invariance holds then
  \begin{equation}
   R^{I}_{\alpha_1 \alpha_2,\beta_1, \beta_2}({\bf p}, {\bf k}) = R^{I}_{\beta_1, \beta_2, \alpha_1 \alpha_2}({\bf p}, {\bf k}) \, ,
   \qquad I= g g, \, q {\bar q} \, .
   \label{CPinv}
  \end{equation}
 The implications of \eqref{bose}, \eqref{CPinv} and of parity invariance on the structure
functions   defined in \eqref{eq:newB}, \eqref{eq:newC} are given in Table~\ref{tab:CPprop}.
In this table the properties of these functions with respect to a naive ${\rm T_N}$ transformation (reversal of momenta and spins)
 and ${\rm CPT_N}$ transformation are collected, too, where just for this purpose, the absorptive part of the respective scattering
  amplitude has been neglected. The entries in  Table~\ref{tab:CPprop} are to be read as follows: For example, for a $T$-invariant
   interaction like QCD one has $b_{n}^{I\pm}(y)=-b_{n}^{I\pm}(y)$ and thus $b_{n}^{I\pm}=0$ at Born level, whereas absorptive parts generated at 
 1-loop render this function non-zero. The above decomposition  of the spin density matrices $R^{I}$ 
  and classification with respect to discrete symmetries is analogous to the analysis  of \cite{Bernreuther:1993hq} where a non-orthogonal basis
  was used.

\begin{table}[h]
 \caption{\label{tab:CPprop} Transformation properties of the structure
functions   defined in \eqref{eq:newB}, \eqref{eq:newC} with respect to  CP, P, and ${\rm T_N}$. The implications of
 Bose symmetry in the  last column apply only
 to $I=gg$. }
\begin{center} 
$$
\begin{array}{|r|r|r|r|r|r|} \hline
\   & \hfil {\rm CP}\hfil   & \hfil {\rm P} \hfil  &\hfil {\rm  T_N}\hfil
&\hfil {\rm  CPT_N}\hfil  &\hfil {\rm Bose}\hfil \\
  \   &  \  &   \
  &\hfil { ({\rm abs}~{\cal T}=0)} \hfil
&\hfil{({\rm abs}~{\cal T}=0)} \hfil  & {\rm symmetry} \\ \hline \hline
A^I(y) & A^I(y) & A^I(y) & A^I(y) & A^I(y) & A^{gg}(-y) \\
b_{r}^{I\pm}(y) & b_{r}^{I\mp}(y) &  -b_{r}^{I\pm}(y) & b_{r}^{I\pm}(y) &
b_{r}^{I\mp}(y) &-b_{r}^{gg\pm}(-y) \\
b_{k}^{I\pm}(y) & b_{k}^{I\mp}(y) & -b_{k}^{I\pm}(y) & b_{k}^{I\pm}(y) &
b_{k}^{I\mp}(y) &b_{k}^{gg\pm}(-y) \\
b_{n}^{I\pm}(y) & b_{n}^{I\mp}(y) & b_{n}^{I\pm}(y) & -b_{n}^{I\pm}(y) &
-b_{n}^{I\mp}(y) &-b_{n}^{gg\pm}(-y) \\
c^I_{rr}(y) & c^I_{rr}(y) & c^I_{rr}(y) & c^I_{rr}(y) &c^I_{rr}(y) & c^{gg}_{rr}(-y)  \\
c^I_{kk}(y) & c^I_{kk}(y) & c^I_{kk}(y) & c^I_{kk}(y) &c^I_{kk}(y) & c^{gg}_{kk}(-y)  \\
c^I_{nn}(y) & c^I_{nn}(y) & c^I_{nn}(y) & c^I_{nn}(y) &c^I_{nn}(y) & c^{gg}_{nn}(-y)  \\
c^I_{rk}(y) & c^I_{rk}(y) & c^I_{rk}(y) & c^I_{rk}(y) &c^I_{rk}(y) & -c^{gg}_{rk}(-y)  \\
c^I_{rn}(y) & c^I_{rn}(y) & -c^I_{rn}(y) & -c^I_{rn}(y) &-c^I_{rn}(y) &c^{gg}_{rn}(-y)  \\
c^I_{kn}(y) & c^I_{kn}(y) & -c^I_{kn}(y) & -c^I_{kn}(y) &-c^{I}_{kn}(y)
&-c^{gg}_{kn}(-y)\\
c^I_{r}(y) & -c^I_{r}(y) & -c^I_{r}(y) & -c^I_{r}(y) &c^I_{r}(y) & -c^{gg}_{r}(-y)  \\
c^I_{k}(y) & -c^I_{k}(y) & -c^I_{k}(y) & -c^I_{k}(y) &c^I_{k}(y) & c^{gg}_{k}(-y)  \\
c^I_{n}(y) & -c^I_{n}(y) & c^I_{n}(y) & c^I_{n}(y) &-c^I_{n}(y) & -c^{gg}_{n}(-y) \\ \hline
\end{array}$$
\end{center}
 \end{table}
 
 The spin density matrices of the $2\to 3$ reactions  that contribute to $\ttbar$ production at NLO in the gauge couplings 
  and are inclusive in $\ttbar$ 
  have the same structure as those discussed above (with coefficient functions which depend on more variables)
   and can also be classified with respect to discrete symmetries.  Table~\ref{tab:CPprop} is useful 
   as a guide for the construction of observables at the level of dileptonic and lepton plus jets final states that  project onto
    contributions to the differential cross section which are even and odd with respect to the above discrete symmetries. These observables
     apply also to investigations beyond LO, cf. section~\ref{sec:obsres}. \\
      Strictly speaking, a classification of observables with respect
  to the discrete symmetries C and CP is not possible in proton proton collisions because the initial $p p$
 state is neither an eigenstate of C nor of CP. Yet, for the main $\ttbar$ production modes, 
 $gg, q{\bar q}\to \ttbar + X$ it can be done. The `contamination' of CP-odd observables by contributions from
   parity-odd mixed QCD weak contributions to 
   $g q ({\bar q}) \to \ttbar q ({\bar q})$ is at the level of a few per mill  \cite{Bernreuther:2010ny}.

 We  add a few remarks on the coefficient functions and on the information contained in Table~\ref{tab:CPprop}.
\begin{itemize}
 \item[i)]    The functions $A^I$,
 $c^I_{nn}$, $c^I_{rr}$, $c^I_{kk}$, and $c^I_{rk}$  are generated by P- and CP-conserving interactions, in particular by QCD
  already at tree level. The functions $A^I$ determine the $\ttbar$ cross section, the $p^t_T$ distribution, etc., while the functions $c^I_{ab}$  
   $(a,b=r,k,n)$  induce P- and CP-even    $\ttbar$ spin correlations.
  \item[ii)] The functions  $b^{\pm}_r$, $b^{\pm}_k$ are generated by P-violating interactions
  and induce longitudinal $t$ and $\bar t$ polarizations (i.e., a polarization in the production plane).
  Non-zero differences $b^{+}_r- b^{-}_r$, $b^{+}_k-b^{-}_k$, which are P-, CP-, and ${\rm  CPT_N}$-odd,
   require CP violation and absorptive parts.
 \item[iii)]
 The coefficient functions  $b^{\pm}_n$  are generated by P- and CP-even absorptive parts in the scattering amplitudes,
  which is the case at 1-loop in the SM, and already at tree level by P-invariant, but CP-violating (effective) interactions.
   If interactions of the latter type are present then $b^{+}_n-b^{-}_n\neq 0$.
     The functions   $b^{\pm}_n$ induce a transverse  $t$ and $\bar t$ polarization (i.e., a polarization normal to the production plane), which would
      differ for $t$ and $\bar t$ if non-standard CP-violating interactions are present.
 \item[iv)]   
Non-zero coefficient functions $c_{kn}$,  $c_{rn}$ require P-odd but CP-even absorptive contributions to the scattering amplitudes. In the SM they
 are generated by the absorptive parts of the 1-loop mixed QCD-weak corrections, and they are very small, see section~\ref{sec:obsres}.
 \item[v)]  
 The coefficient  functions $c_r$,   $c_k$ are generated by  P- and CP-violating interactions. As is well known, in the SM CP violating 
   effects in flavor-diagonal reactions like $a b\to \ttbar$ generated by the Kobayashi-Maskawa phase are tiny. Observable 
   CP effects in hadronic $\ttbar$ production requires sizeable NP CP violation.
 \item[vi)]  
 A non-zero coefficient function $c_n$ requires  P-even but  CP-odd absorptive contributions to the scattering amplitudes.
 In the SM this  coefficient function is tiny. 
\end{itemize}

As an aside we recall that in a time-reversal invariant theory,  expectation values of  ${\rm  T_N}$-odd observables can be generated by
 the absorptive part $i{\cal A}_{fi}={\cal T}_{fi}-{\cal T}_{if}^*$ of the T-matrix element ${\cal T}_{fi}$ of a reaction $i\to f$.
 T-invariance and the unitarity of the $S$ matrix imply
 \begin{equation*}
  |{\cal T}_{fi}|^2 - |{\cal T}_{{\hat f} {\hat i}}|^2=-2 {\rm Im}({\cal A}_{fi}{\cal T}_{fi}^*)- |{\cal A}_{fi}|^2 \, ,
   \end{equation*}
 where ${\hat i}, {\hat f}$ denote the initial and final state with 3-momenta and spins reversed. The left-hand side of this
  equation has the generic form of a ${\rm  T_N}$-odd observable.
  
 \section{Effective NP Lagrangian for hadronic top quark pair production}  
 \label{sec:Lagreff}
 We assume that new physics (NP) effects in hadronic  $\ttbar$ 
 production and decay are induced by new heavy particle exchanges, characterized
 by a mass scale $\Lambda$. Then one may
   construct a local effective Lagrangian ${\cal L}_{NP}$ in terms of the SM degrees of freedom  which  respects the SM (gauge) symmetries
 and describes possible non-resonant new physics interaction structures as long as
  the moduli of the  kinematic invariants of the $\ttbar$ production and decay processes $\lesssim \Lambda$.
  Recent analyses in the context of $\ttbar$ production include
 \cite{AguilarSaavedra:2008zc,AguilarSaavedra:2009mx,Zhang:2010dr,AguilarSaavedra:2010zi,Degrande:2010kt}.
 Here, we follow  \cite{Degrande:2010kt}, but take also CP-violating effective interactions into account.

 \subsection{$\ttbar$ production}
 
 We concentrate on new interactions described by operators of
  mass-dimension ${\rm dim} {\cal O}\leq 6$ which manifest themselves in hadronic
  $\ttbar$ production. We consider only that set of operators which contribute to  $\ttbar$ production through interference with
   the dominant SM amplitudes, i.e., the QCD tree-level amplitudes. \\
   Let's first list the operators which involve gluon fields. We do not consider operators that affect only  the
    pure gluon vertices, such as $GGG$ or $GG{\tilde G}$, where $G$ $({\tilde G})$ denotes the (dual) gluon field strength tensor. For 
     a recent bound on the anomalous coupling associated with $GGG$, see \cite{Buckley:2015nca}. The strength of $GG{\tilde G}$
      is severely constrained by the upper bound on the electric dipole moment of the neutron. For ${\rm dim}{\cal O}\leq 6$ one then remains with the following
      three gauge-invariant operators which involve $t, {\bar t}$ and up to three gluon fields \cite{AguilarSaavedra:2008zc,Zhang:2010dr,Degrande:2010kt}:
   \begin{align}   
  \mathcal O_{gt} &= \bigl[ \bar t_R \gamma^{\mu} T^a D^{\nu} t_R \bigr] \,G^a_{\mu\nu} , \label{eq:Ogt}\\
	      \mathcal O_{gQ} &= \bigl[ \bar Q_L \gamma^{\mu} T^a D^{\nu} Q_L \bigr] \,G^a_{\mu\nu} \label{eq:OgQ},\\
	      \mathcal O_{\text{CDM}} &= \bigl[ \bigl({\tilde\Phi}\bar Q_L \bigr) \sigma^{\mu\nu} T^a t_R \bigr] \,G^a_{\mu\nu} , \label{eq:Hg}.
   \end{align}   
    Here $Q_L=(t_L, b_L)$ is the left-handed third generation doublet and ${\tilde\Phi}=i\sigma_2\Phi^\dagger=(\phi_0^*,-\phi_-)$ is the charge-conjugate
     Higgs doublet field. Moreover, $D_{\mu} = \partial_{\mu} + i g_s T^a G^a_{\mu}$ and 
	 $G^a_{\mu\nu} = \partial_{\mu} G^a_{\nu} - \partial_{\nu} G^a_{\mu} - g_s f^{abc} G^b_{\mu} G^c_{\nu}$ is the gluon field strength tensor.
	  Furthermore,  $T^a$ are the generators of $\text{SU}(3)_c$ in the fundamental representation, with ${\rm tr}(T^aT^b)=\delta_{ab}/2$. \\
	  Using the equation of motion for the gluons one can show \cite{Degrande:2010kt} that the sums $\mathcal O_{gt}+ \mathcal O_{gt}^\dagger$
	   and $\mathcal O_{gQ}+\mathcal O_{gQ}^\dagger$
	   are given by linear combinations of 4-quark operators given below, and are thus redundant if one uses the 4-quark operators below
	    as independent set. On the other hand, the combinations $\mathcal O_{gt}-\mathcal O_{gt}^\dagger$ and $\mathcal O_{gQ}-\mathcal O_{gQ}^\dagger$
	   induce CP-violating interactions, which cannot be expressed in terms of the  the 4-quark operators below. One can therefore use a hermitean effective Lagrangian
	      with these two operator combinations and with \eqref{eq:Hg} that contains three independent complex coefficients. 
	       After spontaneous symmetry breaking,
	       $\langle \Phi\rangle=v/\sqrt{2}$, ($v\simeq 246$ GeV) and restriction to the top-quark gluon sector, one obtains,
	         using operators with definite P and CP properties, real and dimensionless coupling parameters,
	         and the (on-shell) top-quark mass $m_t$ for setting 
	         the mass scale\footnote{Our sign convention for the QCD quark gluon interaction is ${\cal L}_{qg}= -g_s {\bar q}
 \gamma^{\mu} T^a q G^a_\mu$.}:
	       \begin{equation} \label{eq:Lefftg}
	       {\mathcal L}_{\rm NP, g}= -\frac{g_s}{2 m_t}\left[\Hmu {\bar t}\sigma^{\mu\nu}T^a t G^a_{\mu\nu} +
	        \Hd {\bar t}i\sigma^{\mu\nu}\gamma_5 T^a t G^a_{\mu\nu} \right] + \frac{g^2_s}{m^2_t}\left[{\hat c}_{(--)}\mathcal O^g_{(--)} +
	         {\hat c}_{(-+)}\mathcal O^g_{(-+)} \right] \, ,
	        \end{equation}
      where
      \begin{equation}
        \mathcal O^g_{(--)} = \mathcal O_{-} + \mathcal O^{\dagger}_{-}, \qquad 
    	\mathcal O^g_{(-+)} = i\Bigl(\mathcal O_{+} - \mathcal O^{\dagger}_{+}\Bigr),\label{eq:Ogmp}
    	     \end{equation}
 and 
 	\begin{equation}
		\mathcal O_{+} = \bigl[ \bar t \gamma^{\mu} T^a D^{\nu} t \bigr] \,G^a_{\mu\nu},\qquad
		\mathcal O_{-} = i\bigl[ \bar t \gamma^{\mu}\gamma_5 T^a D^{\nu} t \bigr] \,G^a_{\mu\nu} \, .
	\end{equation}
 Here $t$ denotes the top-quark Dirac field. In   \eqref{eq:Lefftg} the real, dimensionless parameters $\Hmu$ and $\Hd$ denote the
  anomalous chromo-magnetic (CMDM) and chromo-electric (CEDM) dipole moments of the top quark in the convention of \cite{Bernreuther:2013aga}.
   Contrary to our previous analysis in \cite{Bernreuther:2013aga} we stay here strictly within the effective field theory approach -- i.e., we do not
    interpret these moments as form factors which may have an absorptive part\footnote{In \cite{Degrande:2010kt}
       only the chromo-magnetic dipole operator was considered. Their parameter $c_{hg}$ is related to $\Hmu$
        by $v{\rm Re}~c_{hg}/(\sqrt{2} \Lambda^2)=-g_s\Hmu/(2 m_t).$}. The
     chromo-magnetic dipole operator is P- and CP-even, $\mathcal  O^g_{(-+)}$ is P-even and CP-odd, while the chromo-electric
      dipole operator and $\mathcal  O^g_{(--)}$ are both P- and CP-odd.
 
    Next, we list the set of gauge-invariant ${\rm dim}{\cal O}= 6$ four-quark operators that yield non-zero tree-level interference terms with the
      QCD amplitude for $q{\bar q}\to \ttbar.$
    We assume universality of the new interactions with respect to the light quarks $q\neq t$.
    Because the respective parton distribution functions suppress $\ttbar$ production by $q{\bar q}$ annihilation 
     for $q=s,c,b$ as compared to $q=u,d$, we take into account only the contributions with $u,d$ in the initial state. \\
      One is then left with the following 7 gauge invariant operators \cite{Zhang:2010dr,Degrande:2010kt}:
    \begin{align}
	      \mathcal O_{Qq}^{(8,\,1)} &= \,\bigl(\bar Q_L \gamma_{\mu} T^a Q_L \bigr)\bigl(\bar q_L \gamma^{\mu} T^a q_L\bigr) ,\label{eq:Qqop1} \\
	      \mathcal O_{Qq}^{(8,\,3)} &= \,\bigl(\bar Q_L \gamma_{\mu} T^a \sigma^A Q_L \bigr)\bigl(\bar q_L \gamma^{\mu} T^a \sigma^A q_L\bigr),\label{eq:Qqop2}  \\
	      \mathcal O_{tu}^{(8)} &= \,\bigl(\bar t_R \gamma_{\mu} T^a t_R \bigr)\bigl(\bar u_R \gamma^{\mu} T^a u_R\bigr), \label{eq:Qqop3} \\
	      \mathcal O_{td}^{(8)} &= \,\bigl(\bar t_R \gamma_{\mu} T^a t_R \bigr)\bigl(\bar d_R \gamma^{\mu} T^a d_R\bigr),\label{eq:Qqop4}  \\
	      \mathcal O_{Qu}^{(8)} &= \,\bigl(\bar Q_L \gamma_{\mu} T^a Q_L \bigr)\bigl(\bar u_R \gamma^{\mu} T^a u_R\bigr), \label{eq:Qqop5} \\
	      \mathcal O_{Qd}^{(8)} &= \,\bigl(\bar Q_L \gamma_{\mu} T^a Q_L \bigr)\bigl(\bar d_R \gamma^{\mu} T^a d_R\bigr), \label{eq:Qqop6} \\
	      \mathcal O_{tq}^{(8)} &= \,\bigl(\bar t_R \gamma_{\mu} T^a t_R \bigr)\bigl(\bar q_L \gamma^{\mu} T^a q_L\bigr) . \label{eq:Qqop7} 
	\end{align}
	Here $q_L=(u_L, u_L)$   denotes the first generation quark doublet, and $\sigma^A$ denote the Pauli matrices
	 (${\rm tr}(\sigma^A \sigma^B)= 2\delta_{AB}.$) \\
	 These operators contribute to the effective NP Lagrangian a term of the form $\sum_{i=1}^7 (c_i/\Lambda^2) \mathcal O_i$,
	  denoting $\mathcal O_{Qq}^{(8,\,1)} =\mathcal O_1$, etc. Again, the terms which involve $b$ quarks are not of interest to
	   us here. 
	  As pointed out in \cite{Degrande:2010kt} it is advantageous to  combine these seven operators such that 
	   one gets four isospin-zero operators with definite P and C properties  and three isospin-one operators. 
	    (In the isospin-one case it is not possible to combine the operators such that they have definite properties
	     with repect to C and P.) In the following $q=(u,d)$ denotes the isospin doublet. One gets 
	     \begin{eqnarray} \label{eq:Leffqua}
	       {\mathcal L}_{\rm NP, q}= {\mathcal L}_{\rm NP, 0} + {\mathcal L}_{\rm NP, 1}  \, ,
	       \end{eqnarray}
	       where the isosoin-zero part is
             \begin{eqnarray} \label{eq:Leffqu0}
	       {\mathcal L}_{\rm NP, 0}=\frac{g_s^2}{2 m_t^2} \sum\limits_{I,J=V,A} {\hat c}_{IJ}  \mathcal O_{IJ} \, .
	       \end{eqnarray}
	       and
	       \begin{eqnarray} \label{eq:Opqu0}
	\mathcal O_{VV} = ({\bar q}\gamma^\mu T^a q)({\bar t}\gamma_\mu T^a t) \, , & \qquad \mathcal O_{AA} = ({\bar q}\gamma^\mu T^a \gamma_5 q)({\bar t}\gamma_\mu \gamma_5 T^a t) \, ,	\\
	\mathcal O_{VA} = ({\bar q}\gamma^\mu T^a  q)({\bar t}\gamma_\mu \gamma_5 T^a t) \, , & \qquad \mathcal O_{AV} = ({\bar q}\gamma^\mu T^a \gamma_5 q)({\bar t}\gamma_\mu T^a t) \, .
	  \end{eqnarray}      	  	 
	   The isospin-one contribution can be represented in the form
	   \begin{eqnarray} \label{eq:Leffqu1}
	       {\mathcal L}_{\rm NP, 1}=\frac{g_s^2}{4 m_t^2} \sum\limits_{i=1}^3 {\hat c}_{i} \mathcal O^1_i \, ,
	       \end{eqnarray}
	       where
	        \begin{eqnarray} \label{eq:Opqu1}
	 \mathcal O^1_1 = & ({\bar q}\gamma^\mu T^a \sigma_3 q)({\bar t}\gamma_\mu T^a t) + ({\bar q}\gamma^\mu \gamma_5 T^a  \sigma_3 q)({\bar t}\gamma_\mu T^a t) \, ,\\
	\mathcal O^1_2 =  & ({\bar q}\gamma^\mu \gamma_5 T^a \sigma_3 q)({\bar t}\gamma_\mu \gamma_5 T^a t) - ({\bar q}\gamma^\mu \gamma_5 T^a  \sigma_3 q)({\bar t}\gamma_\mu T^a t) \, ,\\
	 \mathcal O^1_3 =  & ({\bar q}\gamma^\mu T^a \sigma_3 q)({\bar t}\gamma_\mu \gamma_5 T^a t) + ({\bar q}\gamma^\mu \gamma_5 T^a  \sigma_3 q)({\bar t}\gamma_\mu T^a t) \, .
	 \end{eqnarray}
  Our isospin-zero and -one couplings ${\hat c}_{IJ}, {\hat c}_{i}$ are related to those of \cite{Degrande:2010kt} by
  $g_s^2{\hat c}_{IJ}/m_t^2 = c_{Ij}/\Lambda^2$, $g_s^2{\hat c}_{1}/m_t^2 = c'_{Vv}/\Lambda^2$, $g_s^2{\hat c}_{2}/m_t^2 = c'_{Aa}/\Lambda^2$, 
    $g_s^2{\hat c}_{3}/m_t^2 = c'_{Av}/\Lambda^2$. The relation to the coefficients of the operator basis \eqref{eq:Qqop1} - \eqref{eq:Qqop7} 
     is given in \cite{Degrande:2010kt}. 
     To summarize, we use 
     \begin{equation}\label{eq:LNPsum}
              {\mathcal L}_{\rm NP} = {\mathcal L}_{\rm NP, g} + {\mathcal L}_{\rm NP, q}
       \end{equation}
       The NP contributions to the coefficients of the $\ttbar$ spin density matrices \eqref{Rot} induced by interference
        with the tree-level QCD amplitudes of $gg, q{\bar q}\to \ttbar$ are listed in Appendix~\ref{sec:AppA}.
       
     A side remark concerns $b{\bar b}\to \ttbar$ which receives, apart from the order $\alpha_s^2$ QCD term, a SM contribution of order
      $\alpha_s\alpha$ from the interference of the $t$-channel $W$-exchange -- with possible anomalous $Wtb$ couplings -- 
       and $s$-channel gluon exchange diagram. These SM terms, which are small, are taken into account in section~\ref{sec:obsres}. The  contributions 
        from anomalous $Wtb$ couplings are neglected, because i) they are empirically known to be small (see the next subsection) and ii)
         the respective interference terms are suppressed by the $b\bar b$ parton luminosity.

   \subsection{Top-quark decay}  
   \label{suse:tdecay}
 
 The SM and NP contributions to the parton matrix elements that describe $\ttbar$ 
  production and decay into dileptonic and lepton plus
  jets final states involve also the top-quark decay 
    density matrices for $t\to W b \to \ell\nu b, q {\bar q}' b$.     
   The top-decay vertex $t\to W b$ may also be affected by new physics
 interactions that can also be parametrized by anomalous couplings (cf., for instance,
  \cite{Zhang:2010dr,Zhang:2014rja}).  We use in the following  as  spin-analyzers
  of the (anti)top quark only the charged lepton $\ell^\pm$ from $W^\pm$ decay, and we consider below only lepton angular correlations and
  distributions that are inclusive in the lepton energies. It is known
  \cite{Rindani:2000jg,Grzadkowski:1999iq,Grzadkowski:2002gt,Godbole:2006tq}
 that these observables are not affected by anomalous couplings
 from top-quark decay if these couplings are small, i.e., if a linear approximation is justified.
 This is indeed the case in view of the present upper  bounds  on the
 moduli of these couplings that can be inferred from the measured
 $W$-boson helicity fractions in top-quark decay \cite{Fabbrichesi:2014wva,Cao:2015doa}. In other words, the observables 
  which we analyze in the next section are affected only by possible new physics
   contributions to $\ttbar$ production, which we parametrize  by the anomalous couplings discussed above. \\
   We use the results of \cite{Brandenburg:2002xr} for the polarized semileptonic and non-leptonic top-quark decays in the SM at NLO QCD.
   
  \subsection{Experimental bounds and domain of applicability}  
   \label{suse:rappb} 
   
    The size of the anomalous couplings of the NP interactions described by \eqref{eq:LNPsum} are constrained
     by the experimental data on hadronic $\ttbar$ production which so far show no significant sign of new physics.
      A recent (re)analysis of the contributions
       of only the chromo-moments $\Hmu$ and $\Hd$ to the $\ttbar$ cross section and confrontation with the Tevatron
        and LHC measurements yields correlated bounds. In terms of our  (sign) conventions chosen
         in \eqref{eq:Lefftg},  Ref.~\cite{Hioki:2013hva} determined the allowed region (68\% CL) in the $\Hmu$, $\Hd$ plane
          which may be represented by the bounds  $-0.03\leq\Hmu\leq 0.02$ and $|\Hd|\leq 0.15$.
          Ref.~\cite{Aguilar-Saavedra:2014iga} found a similar result: $-0.046\leq\Hmu\leq 0.024$ and $|\Hd|\leq 0.17$ (95\% CL).
        Ref.~\cite{Franzosi:2015osa} took into account the QCD corrections to  the cross-section contribution of $\Hmu$ and obtained
          $ -0.019\leq\Hmu\leq 0.018$ (95\% CL), putting $\Hd=0$. The CMS experiment analyzed, for dileptonic $\ttbar$ events
           at 7 TeV, the distribution of the difference of the leptonic azimuthal angles, and searched for a 
           contribution of a non-zero $\Hmu$. Using the predictions of  \cite{Bernreuther:2013aga} for interpreting their data, 
           CMS obtained the bound \cite{CMS:2014bea}
            $-0.043 \leq \Hmu\leq 0.117$ (95\% CL). Tighter, but indirect bounds on $\Hd$ and $\Hmu$ were obtained
             using the experimental upper bound on the neutron electric dipole moment \cite{Kamenik:2011dk} 
              and data on rare $B$ meson decays \cite{Martinez:1996cy,Kamenik:2011dk}, respectively.
               (Prospects for correlated bounds on the chromo moments in single top-quark production in the $tW$ mode were recently
                discussed in  \cite{Rindani:2015vya}.)
        
        A global fit to $\ttbar$ data using all dimension-six operators that  are relevant to 
        hadronic top quark production was recently performed in \cite{Buckley:2015nca}. The operator set used in this reference contains also the 
         operators contained in \eqref{eq:LNPsum}, albeit in a slightly different basis. One may use the
          correlated bounds of \cite{Buckley:2015nca} to get order-of-magnitude bounds on the dimensionless couplings of the four-quark
           operators \eqref{eq:Leffqua}, to wit,  $|{\hat c}_{IJ}|, |{\hat c}_{i}| \lesssim 0.4$. 
               
            In the next section we will compute the contributions of the NP interactions \eqref{eq:LNPsum}
    to a number of observables, and we take into account in that section only
     the NP contributions that are linear in the 
    anomalous couplings. As the anomalous couplings correspond to non-renormalizable effective interactions, their
     contributions grow in the regime of large energy-momentum transfer to the $\ttbar$ system. 
      In other words, the linear approximation can be applied with confidence only
   for relatively small anomalous couplings. 
  In order to assess the domain of applicability of this approximation we have considered $\ttbar$ production
   at the LHC (14 TeV). We have computed, for the $\ttbar$ cross section and for a few 
    spin correlation observables
    which will be introduced in the next section, the NP contributions to these
     observables which are linear and quadratic (respectively bilinear)
     in the anomalous couplings. Following Ref. \cite{Degrande:2010kt} we define the range of applicability of the linear approximation 
      by requiring that the contributions linear in the anomalous couplings are  at least twice as large in magnitude as 
       the bilinear ones. Application of this criterion yields $|\Hmu|, |\Hd| \lesssim 0.15$ and $|{\hat c}_{IJ}| \lesssim 0.1$.
      (We did not consider the  contributions of the anomalous couplings ${\hat c}_{i}$ because they are subleading, see
       the next section.) Thus, as far as the chromo-moments are concerned, 
       the  empirical bounds discussed above are  essentially within the domain of applicability of the linear approximation.
   
\section{Observables and results}
\label{sec:obsres}

In this section we consider $\ttbar$ production at the LHC for center-of-mass energies 8, 13, and 14 TeV. 
 First we determine the SM contributions and those of ${\mathcal L}_{\rm NP}$ to the $\ttbar$ cross section $\sigma_{\ttbar}$ and
  the  $\ttbar$ charge asymmetry $A_C$. Then we introduce, for dileptonic and lepton plus jets final states,  a set of correlation
   and polarization observables and compute their distributions. 
   
  As far as the SM interactions are concerned we compute the four P-even and CP-even correlation observables that will be defined
   in section~\ref{sec:corrllnew} at NLO QCD (order $\alpha_s^3$) \cite{Bernreuther:2004jv} including the weak and mixed QCD-weak corrections of order
    $\alpha^2, \alpha_s\alpha$, $\alpha_s^2\alpha$, $\alpha_s\alpha^2$ \cite{Bernreuther:2005is,Bernreuther:2006vg,Bernreuther:2008md,Bernreuther:2010ny,Bernreuther:2013aga}. 
    We label these SM contributions by the acronym  `NLOW'.
    The mixed QCD-QED contributions to these spin correlation observables have not yet been calculated. We expect them to be small, of the 
     order of a few percent of the QCD corrections. The combinations of  polarization observables 
   considered in section~\ref{sec:corrllnew} are, apart from one combination,  P-odd and/or CP-odd. 
    These observables receive no QCD or QED contributions. Only one of these polarization observables,  
     the sum of the transverse $t$ and $\bar t$ polarization, receives, at the order of perturbation theory considered here,
     QCD contributions of order $\alpha_s^3$,  and  mixed QCD-weak and mixed QCD-QED contributions of order $\alpha_s^2\alpha$, which
      are taken  into account. (Contributions of order $\alpha^2\alpha_s$ to the transverse polarization
       are very small and are not considered here.) The label `EW' refers to the electroweak contributions.
      Our computational set-up is described in \cite{Bernreuther:2010ny}. \\
    
      In the SM the total  $\sigma_{\ttbar}$ cross section is known to NNLO QCD (order  $\alpha_s^4$) including soft-gluon resummation
       at next-to-next-to-leading logarithmic (NNLL) order  \cite{Czakon:2013goa}. 
        The mixed QCD-weak corrections to the cross section (NLOW) were computed in \cite{Bernreuther:2005is,Bernreuther:2006vg,Bernreuther:2008md,Kuhn:2005it,Kuhn:2006vh}
        and the mixed QCD-QED corrections were determined in  \cite{Hollik:2007sw}. The NLOW corrections to the LHC $\ttbar$ cross section
         are negative, about $\sim -2\%$ of the LO QCD cross section, while the QCD-QED corrections are positive,  about $\sim 1\%$ according 
          to  \cite{Hollik:2007sw}.
    The LHC charge asymmetry $A_C$ defined below is computed at NLO QCD, including the mixed  QCD-weak  and mixed   QCD-QED corrections
    \cite{Bernreuther:2012sx}. These mixed EW corrections are known to be important for $A_C$.
    The computation of $A_C$ to this order of perturbation theory, which is labeled `NLO+EW',  is present state-of-the-art. 
     The Tevatron charge asymmetry was recently computed to NNLO accuracy \cite{Czakon:2014xsa}.

    Furthermore, we compute the  interferences of the NP amplitudes induced by ${\mathcal L}_{\rm NP}$
   and the tree-level QCD amplitudes, but keep only the terms linear in the anomalous couplings.   
   We employ the following input parameters: For the top-quark mass in the on-shell scheme we use
 $m_t=  173.34 \ {\rm GeV}.$  Moreover, we use $\Gamma_t  =1.3  \ {\rm GeV}$, $m_Z=91.2$ GeV,   $m_W=  80.4 \ {\rm GeV}, 
 \Gamma_W = 2.09 \ {\rm GeV}$, $m_H=125$ GeV, $\alpha(m_t)=0.008$, and we use the CT10 NLO parton distribution functions (PDF) \cite{Lai:2010vv}. 
 This PDF set provides also the QCD coupling $\alpha_s$ in the $\overline{\rm MS}$ scheme.
  For the renormalization and factorization scale, we use $\mu_R = \mu_F \equiv \mu$, $\mu=m_t$, and estimate scale uncertainties by varying $\mu$
   between $m_t/2$ and $2 m_t$.

\subsection{Total $\ttbar$ cross section and charge asymmetry at the LHC}
 \label{sec:LHCxsec}
 
 Our results for $\sigma_{\ttbar}$ are listed in table~\ref{tab:sigmAC}. The uncertainties refer to the scale uncertainties.  
 In the linear approximation, only the P- and CP-even operators contained in ${\mathcal L}_{\rm NP}$ contribute; the CMDM operator, whose contribution is
  largest, the isospin-zero operator  ${\mathcal O}_{VV}$, and the P-even part of the isospin-one operator  ${\mathcal O}^1_1$, whose contribution is smallest.
   The smallness of the isospin-1 compared to the isospin-0 contribution to $\sigma_{\ttbar}$, to the charge asymmetry defined below,  and to the observables
    of the next section is due to the fact that a sizeable fraction of these
    contributions comes from the kinematic region near threshold. In this region, the contributions of $u{\bar u}\to \ttbar$  and $d{\bar d}\to \ttbar$  are of the same order.
   The SM result for the cross section listed in table~\ref{tab:sigmAC} is given here only for reference purposes. 
     Comparing experimental results to predictions one should use the most precise available 
      SM result  on $\sigma_{\ttbar}$.  This is the result of \cite{Czakon:2013goa} mentioned above. Rather than performing a three-parameter
   fit to cross section results at the Tevatron and the LHC (7 and 8 TeV) which is left to a future analysis, we make here a crude estimate of the bound on $\hat c_{VV}$ by putting the
    anomalous couplings $\Hmu$ and $\hat c_1$ to zero. Using, for 8 TeV,  $\sigma_{\rm exp.}=241.5\pm 8.5$ pb \cite{ATLAS:2014aaa}
     and  the NNLO+NNLL QCD prediction  \cite{Czakon:2013goa,Czakon:2011xx}   $\sigma_{\rm th}=253 \pm 14$ pb and neglecting the electroweak contributions, we get 
      the 95 $\%$ C.L. bound $-0.18  \leq \hat c_{VV} \leq 0.09$. The lower bound is outside of the condition $|{\hat c}_{VV}| \lesssim 0.1$ 
       for the linear approximation  derived in section~\ref{suse:rappb} if the other anomalous couplings are put to zero.

	\begin{table}[h]
		\centering
		\caption{\label{tab:sigmAC} The SM and NP contributions to the $\ttbar$ cross section and the LHC charge asymmetry \eqref{eq:LHCchaas}.}
		
		\begin{tabular}{|c||c|c|c|c|c|}\hline
		\multicolumn{2}{|c|}{} & NLOW & $\propto \hat c_{VV}$ & $\propto \hat c_1$ & $\propto \Hmu$ \\\hline\hline
		\multirow{3}{*}{$\sigma$[pb]} 	& 8 TeV   &$207.78^{+23.51}_{-25.95}$ & $246.23^{+49.08}_{-67.10}$ & $32.33^{+6.13}_{-8.23}$ & $761.03^{+255.18}_{-178.27}$\\
						& 13 TeV  & $687.32^{+78.35}_{-80.23}$ & $581.16^{+104.82}_{-138.98}$ & $71.06^{+12.23}_{-15.90}$ & $2490.91^{+723.94}_{-527.16}$\\
						& 14 TeV  & $814.38^{+92.89}_{-94.02}$ & $654.65^{+116.20}_{-153.50}$ & $79.14^{+13.37}_{-17.40}$ & $2945.65^{+836.38}_{-612.79}$\\\hline
		\multicolumn{2}{|c|}{  } 	& NLO + EW & $\propto \hat c_{AA}$  & \multicolumn{2}{|c|}{$\propto \hat c_{2}$} \\\hline\hline
		\multirow{3}{*}{$A_C$}		& 8 TeV  & $(1.11^{+0.04}_{-0.04})\cdot 10^{-2}$  & $0.446^{+0.017}_{-0.017}$  & \multicolumn{2}{|c|}{$(9.26^{+0.42}_{-0.39})\cdot 10^{-2}$}\\
						& 13 TeV & $ (0.75^{+0.04}_{-0.05})\cdot 10^{-2}$   & $ 0.341^{+0.011}_{-0.011}$  & \multicolumn{2}{|c|}{$(7.07^{+0.27}_{-0.27})\cdot 10^{-2}$}\\
						& 14 TeV & $ (0.66^{+0.05}_{-0.04})\cdot 10^{-2}$        & $0.327^{+0.010}_{-0.010}$  & \multicolumn{2}{|c|}{$(6.77^{+0.25}_{-0.25})\cdot 10^{-2}$}\\\hline
		
		\end{tabular}
	\end{table}%
	
     Table~\ref{tab:sigmAC} contains also our SM and NP  prediction for the $\ttbar$ charge asymmetry at the LHC which is defined by
       \begin{equation} \label{eq:LHCchaas}
     A_C = \frac{ \sigma(\Delta|y|>0) - \sigma(\Delta|y|<0)}{ \sigma(\Delta|y|>0) + \sigma(\Delta|y|<0)}  
    \end{equation}
     where $\Delta|y|=|y_t|-|y_{\bar t}|$ is the difference of the moduli of the $t$ and $\bar t$ rapidities
 in the laboratory frame.  Our SM predictions, labeled `NLO + EW',   contain apart from the 
  mixed QCD-weak interaction corrections also the mixed QCD-QED corrections of order $\alpha_s^2\alpha$ \cite{Bernreuther:2012sx}.  
  We have computed $A_C$ by expanding the ratio \eqref{eq:LHCchaas}, see Eq. \eqref{eq:Cexp} below.
   The NLOW predictions in  table~\ref{tab:sigmAC} are in accord with the results of \cite{Bernreuther:2012sx,Kuhn:2011ri} where  different
  PDF sets were used.                                 
 The NP contributions result from the  isospin-zero operator  ${\mathcal O}_{AA}$ and the P-even part of
  the isospin-one operator  ${\mathcal O}^1_2$. These operators induce contributions to the differential cross section that are odd
   under interchange of the $t$ and $\bar t$ momenta while those of the initial (anti)quark are kept fixed. Using the recent CMS measurement \cite{Khachatryan:2015oga}
    with their 8 TeV
    data, $A_C=0.001\pm 0.008$, and putting the coupling $\hat c_2$ to zero, we obtain
      the 95~$\%$ C.L. bound $-0.06  \leq \hat c_{AA} \leq 0.01$; cf. also \cite{Khachatryan:2015oga,Gabrielli:2011zw}. This result justifies a posteriori
       the use of the linear approximation. 
      New physics contributions to the  $\ttbar$ charge asymmetry $A_{FB}$ at the Tevatron
       were analyzed in terms of the above effective interactions in 
       \cite{Degrande:2010kt,Gabrielli:2011jf,AguilarSaavedra:2011vw,Delaunay:2011gv}.
      The latest experimental results on the Tevatron asymmetry $A_{FB}$ are somewhat larger -- but compatible -- with the
       most precise SM results \cite{Czakon:2014xsa}. This implies $\hat c_{AA}>0$ and $\hat c_{AA}$ is bounded from above.
      The interaction described by ${\mathcal O}_{AA}$ is the effective field theory version of a number of
       NP models which were invoked to resolve  the seeming discrepancy between the experimental results on the  Tevatron $A_{FB}$ and the
         SM predictions. These models include the axigluon model and flavor-changing $t$-channel $Z'$ exchange in $u{\bar u}\to \ttbar$; see
         \cite{Aguilar-Saavedra:2014kpa} for a recent overview
      of the NP models discussed in the context of the $\ttbar$ charge asymmetry.
      
      The linear approximation for the contributions of the NP interactions to the cross section and the charge
       asymmetry is justified if $|\Hmu|< |{\hat\mu}_{t,\rm max}|\simeq  0.15$ and $|{\hat c}_{IJ}| <  |{\hat c}_{IJ, \rm max}| \simeq 0.1$, see section~\ref{suse:rappb}.
       We assume in the following that the latter bound holds also for anomalous couplings ${\hat c}_{i}$.  
       With these bounds we may make a crude estimate of the maximal difference between (future) experimental measurements of an observable and the SM prediction that
        will still allow the the use of the linear approximation. In the case of the $\ttbar$ cross section, we estimate this maximal difference by
        \begin{equation}\label{eq:maxdiff}
         \delta \sigma_{\ttbar}=|\sigma_{\ttbar}^{\rm exp.}- \sigma_{\ttbar}^{\rm SM}|_{\rm max} =|a_1\hat c_{VV,\rm max}|
                                      +| a_2 \hat c_{1,\rm max}| + | a_3 {\hat\mu}_{t,\rm max}| \, ,
                  \end{equation}
         where the coefficients $a_i$ are given in Table~\ref{tab:sigmAC}. (We take the values of the $a_i$
          for $\mu=m_t$.) The maximal difference  $\delta A_C$ is determined in analogous fashion. 
       In Table~\ref{tab:maxdiff1} our estimates of $\delta \sigma_{\ttbar}$ and $\delta A_C$ are given for the LHC (13 TeV). As these are crude estimates  we have refrained
        from taking the scale uncertainties of the $a_i$ into account. For the other LHC center-of-mass energies, the 
          $\delta \sigma_{\ttbar}$ and $\delta A_C$ are obtained in analogous fashion using the coefficients of  Table~\ref{tab:sigmAC}. 
       
  \begin{table}[h]
	\caption{Moduli of maximal differences between future experimental result and SM prediction, for the LHC (13 TeV), for which the 
	linear approximation can be applied.}
	\centering
  	\begin{tabular}{|c||c|c|}\hline
  		Observable & no cut & $m_{t\bar t} < 500 \text{GeV}$ \\\hline\hline
  		$\delta\sigma_{\ttbar}$[pb] & 440 & 242  \\\hline
  		$\delta A_C$ & 4.1$\times 10^{-2}$ & 1.7$\times 10^{-2}$ \\\hline
  	\end{tabular}
  	 \label{tab:maxdiff1}
  \end{table}

  	Table~\ref{tab:sigmAClowc} contains our  predictions for $\sigma_{\ttbar}$ and $A_C$ for events with $\ttbar$ invariant mass  $\mttbar \leq$ 500 GeV.
  	The mixed QCD-QED contributions to  $\sigma_{\ttbar}(\mttbar \leq 500~{\rm GeV})$, which are not included in this table, add $\sim 1\%$ to the cross-section values
  	 given in this table. The NNLO QCD corrections are not yet available.
  	  For $A_C$, the ratio of the isospin-one and isospin-zero NP contributions is approximately the same as in the inclusive case listed in table~\ref{tab:sigmAC}.
  	 In order to determine constraints on both anomalous couplings $\hat c_{AA}$ and  $\hat c_2$ from data on the LHC charge asymmetry alone,
  	  one may compute the differential charge asymmetry in narrower bins of $\mttbar$, or bins of the rapidity and transverse 
  	  momentum of the $\ttbar$ system. \\
  	  In analogy to \eqref{eq:maxdiff} one may estimate the maximal differences $\delta \sigma_{\ttbar}(\mttbar \leq 500~{\rm GeV})$ and \\
  	  $\delta A_C(\mttbar \leq 500~{\rm GeV})$ that would allow the use of the linear approximation. They are given in Table~\ref{tab:maxdiff1}.

  	      In the region where, for instance, the $\ttbar$ invariant mass $\mttbar$ is large, the linear approximation in the NP couplings 
  	    can  no longer be justified just by assuming that the anomalous couplings are small, see section~\ref{suse:rappb}.
  	   While it is not difficult to compute $\sigma_{\ttbar}$ and $A_C$ beyond this approximation, the following complication in the comparison with data 
  	    arises: there is now a proliferation of contributions (being odd
  	    in the scattering angle $y$ defined in section~\ref{sec:ortho}) which involve, apart from $\hat c_{AA}$ and  $\hat c_2$, 
  	    also the other anomalous couplings.
  	    Information on both $\hat c_{AA}$ and $\hat c_2$ can be obtained either from considering $A_C$ at different c.m. energies 
  	    and the Tevatron asymmetry simultaneously,
  	     or from $A_C$ and  (future) measurements of the inclusive $\ttbar$ spin correlation and polarization observables defined in the next section.

	\begin{table}[h]
		\centering
		\caption{ \label{tab:sigmAClowc} The SM and NP contributions to the $\ttbar$ cross section and the LHC charge asymmetry \eqref{eq:LHCchaas}
		 events with $\mttbar \leq$ 500 GeV.}
		\begin{tabular}{|c||c|c|c|c|c|}\hline
		\multicolumn{2}{|c|}{--} 			& NLOW 		& $\propto \hat c_{VV}$ 	& $\propto \hat c_1$ 			& $\propto \Hmu$ \\\hline\hline
		\multirow{3}{*}{$\sigma$[pb]} 	& 8 TeV  	& $129.73^{+15.81}_{-16.19}$	& $104.67^{+19.16}_{-25.43}$ 	& $12.89^{+2.22}_{-2.89}$ 		& $480.35^{+152.01}_{-108.09}$\\
						& 13 TeV  	& $394.22^{+47.75}_{-45.95}$	& $212.03^{+34.11}_{-43.75}$ 	& $23.98^{+3.61}_{-4.58}$ 		& $1455.30^{+393.45}_{-292.17}$\\
						& 14 TeV  	& $461.06^{+55.77}_{-53.02}$		& $234.11^{+36.95}_{-47.01}$ 	& $26.12^{+3.88}_{-4.86}$ 		& $1702.52^{+448.04}_{-335.37}$\\\hline
		\multicolumn{2}{|c|}{--} 			& NLO + EW 		& $\propto \hat c_{AA}$  	& \multicolumn{2}{|c|}{$\propto \hat c_{2}$} \\\hline\hline
		\multirow{3}{*}{$A_C$}		& 8 TeV  	&  $ (0.96^{+0.04}_{-0.04})\cdot 10^{-2}$ 		& $0.209^{+0.011}_{-0.011}$  	& \multicolumn{2}{|c|}{$(4.38^{+0.25}_{-0.24})\cdot 10^{-2}$}\\
						& 13 TeV 	&  $ (0.63^{+0.04}_{-0.04})\cdot 10^{-2}$ 		& $0.141^{+0.007}_{-0.007}$  	& \multicolumn{2}{|c|}{$(2.89^{+0.16}_{-0.15})\cdot 10^{-2}$}\\
						& 14 TeV 	&  $ (0.50^{+0.04}_{-0.04})\cdot 10^{-2}$ 		& $0.133^{+0.006}_{-0.006}$  	& \multicolumn{2}{|c|}{$(2.71^{+0.14}_{-0.14})\cdot 10^{-2}$}\\\hline
		
		\end{tabular}
	\end{table}%

\subsection{Angular distributions and correlations for dilepton and $\ell +$ jets events}
\label{sec:corrllnew}

We analyze $\ttbar$ production  at the LHC and subsequent decay into  dileptonic
final states, 
\begin{equation}\label{ppdilep}
p p \to t + {\bar t} + X\to \ell^ + \ell'^-  + \ \text{jets}  + E_T^{\rm miss}\, ,
\end{equation}
 and into  lepton plus jets final states,
\begin{eqnarray}
p p & \to t + {\bar t} + X & \to \ell^+ +  \ \text{jets} + E_T^{\rm miss}\, , \label{ppljt}\\
p p & \to t + {\bar t} + X& \to \ell^- + \ \text{jets} + E_T^{\rm miss} \, .\label{ppljtbar}
\end{eqnarray} 
where $\ell,\ell'= e,\mu,\tau.$

For the dileptonic events \eqref{ppdilep} we consider 
four-fold dilepton angular distributions. Without acceptance cuts, besides cuts on the $\ttbar$ invariant mass,
 these normalized distributions  have the form 
\begin{align}
	\frac{1}{\sigma} \frac{d\sigma}{d\Omega_+ d\Omega_-} = \frac{1}{(4\pi)^2}\Bigl( 1 + {\bf B}'_1\cdot \hlp  + {\bf B}'_2 \cdot \hlm - 
 \hlp \cdot C^{\prime} \cdot \hlm \Bigr), \label{eq:ll4fold}
\end{align}
where $d\Omega=d\cos\theta d\phi$. The unit vectors $\hlp$, $\hlm$ are the $\ell^+$ and $\ell'^-$ directions of flight in the $t$ and $\bar t$ rest frames, respectively, which are
 reached from the $\ttbar$ ZMF by rotation-free boosts. The coefficients, i.e., the vectors   ${\bf B}'_1$,  ${\bf B}'_2$ and the matrix $C'$
  contain the information from the $\ttbar$ production and the $t$ and $\bar t$ decay density matrices, which may be computed in the SM or in a NP model.
 As already mentioned in section~\ref{suse:tdecay},
  we consider the main SM top-decay mode $t\to W b$. We parametrize $\hlp, \hlm$ in terms of spherical coordinates, 
 but we allow for different polar axes, i.e.,  different reference axes $\ha$ and $\hb$.
  We define
\begin{align}
	\cos\theta_{+} = \hlp \cdot \ha \, , \qquad 	\cos\theta_{-} =\hlm \cdot \hb \, .
\end{align}
Integrating over the azimuthal angles we obtain the polar angle double distributions for a choice of reference axes $\ha, \, \hb$:
\begin{align}
	\frac{1}{\sigma} \frac{d\sigma}{d\cos\theta_+ d\cos\theta_-} = \frac{1}{4}\Bigl( 1 + B_1 \, \cos\theta_+ +  B_2 \, \cos\theta_- - C\, \cos\theta_+\cos\theta_- \, \Bigr).
	\label{eq:doublediff}
\end{align}
The sign in front of $C$ is chosen such that in the absence of acceptance cuts \cite{Bernreuther:2001rq,Bernreuther:2004jv}:
\begin{equation}\label{eq:CDspinas}
 C(\ha, \hb) = \kappa_\ell^2 \;  
\frac{\sigma(\uparrow \uparrow)+\sigma(\downarrow \downarrow)
  - \sigma(\uparrow \downarrow)- \sigma(\downarrow \uparrow)}{\sigma(\uparrow \uparrow)+\sigma(\downarrow \downarrow)
  + \sigma(\uparrow \downarrow)+ \sigma(\downarrow \uparrow) } \,
 \end{equation}
 where the ratio on the right-hand side of this formula is the double spin-asymmetry at the level of the $t$, $\bar t$ intermediate states.
  (The first (second) arrow refers to the spin state of the $t$ $({\bar t})$ quark with respect to the axis $\ha$ $(\hb)$.)
 The top-spin analyzing power of the charged
 lepton from top decay is denoted by $\kappa_\ell$. (In the convention used here,
 $\kappa_{\ell^+}=\kappa_{\ell^-}=\kappa_\ell.$) At NLO QCD its value is $\kappa_\ell=0.999$ \cite{Brandenburg:2002xr}. \\
 The polarization degrees of the ensembles of $t$ and $\bar t$ quarks in $\ttbar$ events 
  with respect  to the reference axes  $\ha, \, \hb$  are given by
\begin{equation} \label{eq:polde}
P(\ha)=\langle 2 \Sp\cdot\ha\rangle \, , 
\qquad {\overline P}(\hb)=\langle 2 \Sm\cdot\hb\rangle \, ,
\end{equation}
 where $\Sp$, $\Sm$ denotes the $t$ and $\bar t$ spin operator,
respectively. They are related to $B_{1,2}$ by
\begin{equation} \label{c4:relpol}
  B_1(\ha) = P(\ha)~\kappa_\ell \, , \qquad B_2(\hb) = - {\overline P}(\hb)~\kappa_\ell \, .
\end{equation}
 That is, the relative signs  in front of the $B_{1,2}$ in  \eqref{eq:doublediff}  are chosen such that in a CP-invariant theory and for 
  the choice $\ha  = - \hb$  we have
\begin{align}
	B_1 = B_2 \, . \label{B1B2}
\end{align}

Our  choice of reference axes $\ha$ and $\hb$ which we use here is adapted to the
 orthonormal basis at the parton level 
 introduced in section \ref{sec:ortho}, but is not identical to it. As in 
section \ref{sec:ortho}, we use the unit vector $\hk$ which is the top quark direction
 of flight in the $\ttbar$ ZMF. Moreover, we use the direction of  one of the proton beams in the
  laboratory frame, ${\hp}_p$, and define unit vectors  ${\hr}_p$ and ${\hn}_p$ as follows: 
 \begin{align}
  {\hp}_p =(0,0,1) \, , & \quad {\hr}_p =\frac{1}{r_p}({\hp}_p -y_p \hk) \, , & \quad {\hn}_p =\frac{1}{r_p} ({\hp}_p \times \hk) \, , 
   \label{proONB} \\
    & \quad y_p = {\hp}_p\cdot \hk \, , & \quad r_p=\sqrt{1-y^2_p}  \, . \nonumber
   \end{align}
Only in the case of $2\to 2$ parton reactions and if the incoming parton 1 is parallel to  ${\hp}_p$, the unit vectors  defined in \eqref{orhtoset}
 are the same as those in \eqref{proONB}. With the set \eqref{proONB} we define 
  the reference axes $\ha$ and $\hb$ listed in table~\ref{tab:axes}. 
 The factors $\text{sign}(y_p)$ are required because
of the Bose symmetry of the initial $gg$ state. 

Instead of extracting the spin correlation coefficients $C$ (for a choice of axes) from the double distributions  (\ref{eq:doublediff}) 
one may analyze the 1-dimensional distributions of the
 observables
\begin{align}
\xi = \cos\theta_+\cos\theta_- \, . \label{obsxi}
\end{align}
Labels are suppressed here, i.e., we consider in the following $\xi_{nn}, \xi_{rr},  \xi_{kk},$ $\xi_{rk}, \xi_{rn},$ etc. The 
 one-dimensional distribution is obtained in standard fashion by inserting 
\begin{align}
    1 = \int d\xi \delta(\xi - \cos\theta_+\cos\theta_-)
\end{align}
 into the integral $\int d\cos\theta_+ d\cos\theta_-$ over the double distribution (\ref{eq:doublediff}). One arrives at 
\begin{align}
	\frac{1}{\sigma} \frac{d\sigma}{d\xi} = \frac{1}{2}\Bigl( 1 - C\xi \Bigr)\,\ln\bigl(\frac{1}{|\xi|}\bigr) \, ,
	\label{eq:singlediff2}
\end{align}
 which may be used for fits to unfolded experimental results. We recall that
  \begin{align} \label{eq:Cxiform}
               C = - 9 \: \langle \xi \rangle \, .
\end{align} 
 
  In this paper we apply the distributions \eqref{eq:doublediff} and \eqref{eq:singlediff2} only to dileptonic final states.
 Integrating out 
one of the cosines in the double distribution (\ref{eq:doublediff}) one gets
\begin{align}
	\frac{1}{\sigma} \frac{d\sigma}{d\cos\theta_{\pm}} = \frac{1}{2}\Bigl( 1 + B_{1,2} \, \cos\theta_{\pm}  \Bigr).
	\label{eq:singlediff1}
\end{align}
The slope of this distribution, $B_1 (B_2)$, measures the polarization of $t$ $({\bar t})$ quarks with respect to 
the chosen reference  axis. One may also compute/measure these polarizations 
  using the relation  $B_{1,2} = 3 \: \langle \cos\theta_\pm \rangle .$
  If no acceptance cuts are applied, the  coefficient $B_{1}$ $(B_2)$ is 
  the same both for the distribution of $\ell^+$ $(\ell^-)$ from dilepton and from lepton plus jets events.

  From now on we use the following notation. The label  $(a,\, b)$ refers   to the choice of reference axes
 $\ha$ and $\hb$ from  table~\ref{tab:axes}. The correlation coefficient   $C(a, \, b)$ in (\ref{eq:doublediff}) or 	\eqref{eq:singlediff2} 
 is associated with this choice of axes, and likewise $B_1(a)$    and   $B_2(b)$.

  In addition to  \eqref{eq:singlediff1}, we compute also the top and antitop polarization 
  with respect to another set of reference axes:
 \begin{equation} \label{eq:sumBkrstar}
  B_1(k^*), \quad  B_2(k^*) \, , \qquad B_1(r^*), \quad  B_2(r^*)   \, ,
\end{equation} 
where the labels $k^*$ and $r^*$ refer to the following vectors: 
\begin{align}
 k^*: \quad & {\ha} = {\rm sign}(\Delta|y|)~{\hk} \, , \quad {\hb} = -{\rm sign}(\Delta|y|)~{\hk}\, , \label{eq:kstardef}\\
 r^*: \quad & {\ha} = {\rm sign}(\Delta|y|)~\text{sign}(y_p)~{\hr}_p \, ,\quad   {\hb} = -{\rm sign}(\Delta|y|)~\text{sign}(y_p)~{\hr}_p \, . \label{eq:rstardef}
\end{align}
Here  $\Delta|y|=|y_t|-|y_{\bar t}|$ denotes the difference of the moduli of the $t$ and $\bar t$ rapidities
 in the laboratory frame. The reason for introducing these additional polarization observables is that they 
 are sensitive to the  NP contributions
   from the operator   $\mathcal O_{AV}$ and from a P-odd combination of the operators  $\mathcal O^1_i$ 
    (defined in \eqref{eq:Opqu0} and \eqref{eq:Opqu1}, respectively), while 
    $B_{1,2}(k), B_{1,2}(r)$ project onto the contributions of $\mathcal O_{VA}$ and  $\mathcal O^1_3$.

\begin{table}[!t]
	\caption{Choice of reference axes. The unit vectors ${\hn}_p$, ${\hr}_p$ and the 
	variable $y_p$ are defined in \eqref{proONB}.}
	\label{tab:axes}
	
	\centering
    \begin{tabular}{|c c|c|c|} \hline
	& label    	&	~~~$\ha $ ~~~		&	~~~$\hb$~~~ \\ \hline
transverse &	n		&	~~~$\text{sign}(y_p)\; {\hn}_p$~~~	&	~~~$-\text{sign}(y_p)\; {\hn}_p$~~~ \\
r axis & 	r		&	$\text{sign}(y_p)\; {\hr}_p$	&	$-\text{sign}(y_p)\; {\hr}_p$ \\
helicity &	k		&	$\hk$		&	$-\hk$ \\ \hline
	\end{tabular}
\end{table}

The  coefficients $C(a, \, b)$, $B_1(a)$,   and   $B_2(b)$ (and sums and differences) are listed in  table~\ref{tab:C-b-coeff}. 
 They can be measured with the  2-dimensional distributions \eqref{eq:doublediff} or with the 1-dimensional
  distributions of $\xi_{ab}$,  respectively with \eqref{eq:singlediff1}.
  Column 2 of this table indicates which 
 coefficient of the $ij\to \ttbar X$ production density matrices introduced in section~\ref{sec:ortho} contributes to the respective
  correlation coefficient. The type of interactions, i.e., the type of  contributions in
   the differential parton cross sections to which a  correlation is sensitive, is given in column 3.

\begin{table}[t]
	\caption{The correlation coefficients of (\ref{eq:doublediff}), respectively of  \eqref{eq:Cxiform}, \eqref{eq:singlediff1} and \eqref{eq:sumBkrstar},
         and sums and differences  for different choices of reference axes.}
	\label{tab:C-b-coeff}
	
	\centering
	 \begin{tabular}{|c|c|c|} \hline
		Correlation		&	                	&	sensitive to \\\hline\hline
		$C(n,\,n)$ 		&	$c^I_{nn}$		&	P-, CP-even\\
		$C(r,\,r)$ 		&	$c^I_{rr}$		&	P-, CP-even\\
		$C(k,\,k)$ 		&	$c^I_{kk}$		&	P-, CP-even\\
		$C(r,\,k)+C(k,\,r)$ 	&	$c^I_{rk}$		&	P-, CP-even\\
		$C(n,\,r)+C(r,\,n)$ 	&	$c^I_{rn}$                &       P-odd, CP-even, absorptive\\
		$C(n,\,k)+C(k,\,n)$ 	&	$c^I_{kn}$		&	 P-odd, CP-even,  absorptive\\
		$C(r,\,k)-C(k,\,r)$ 	&	$c^{I}_n$	&	 P-even, CP-odd, absorptive \\
		$C(n,\,r)-C(r,\,n)$ 	&	$c^{I}_k$	&	 P-odd, CP-odd\\
		$C(n,\,k)-C(k,\,n)$ 	&	$-c^{I}_r$	&	 P-odd, CP-odd\\
		$B_1(n) + B_2(n)$ 	&	$b^{I+}_{n} + b^{I-}_{n}$	&	P-, CP-even, absorptive\\
		$B_1(n) - B_2(n)$ 	&	$b^{I+}_{n} - b^{I-}_{n}$ 	&	 P-even, CP-odd\\
		$B_1(r) + B_2(r)$ 	&	$b^{I+}_{r} + b^{I-}_{r}$ 	&	 P-odd, CP-even \\
		$B_1(r) - B_2(r)$ 	&	$b^{I+}_{r} - b^{I-}_{r}$ 	&	 P-odd, CP-odd, absorptive\\
		$B_1(k) + B_2(k)$ 	&	$b^{I+}_{k} + b^{I-}_{k}$ 	&	 P-odd,CP-even \\
		$B_1(k) - B_2(k)$ 	&	$b^{I+}_{k} - b^{I-}_{k}$ 	&	 P-odd, CP-odd, absorptive\\
	        $B_1(k^*) + B_2(k^*)$ 	&	$b^{I+}_{k} + b^{I-}_{k}$ 	&	 P-odd,CP-even \\
		$B_1(k^*) - B_2(k^*)$ 	&	$b^{I+}_{k} - b^{I-}_{k}$ 	&	 P-odd, CP-odd, absorptive\\
		$B_1(r^*) + B_2(r^*)$ 	&	$b^{I+}_{r} + b^{I-}_{r}$ 	&	 P-odd, CP-even \\
		$B_1(r^*) - B_2(r^*)$ 	&	$b^{I+}_{r} - b^{I-}_{r}$ 	&	 P-odd, CP-odd, absorptive	\\ \hline	
		\end{tabular}
\end{table}

\noindent
Tables~\ref{tab:Ccoefnum},~\ref{Bcoefnum} contain our results for the inclusive correlation coefficients 
$C(a, \, b)$, $B_1(a)$, $B_2(b)$ and   $B_1(a^*)$, $B_2(b^*)$
 for the LHC at 8, 13, and 14 TeV, 
  while the coefficients in tables~\ref{tab:Ccoefnumlowc},~\ref{Bcoefnumlowc} 
 refer to events with $\ttbar$ invariant mass  $\mttbar \leq$ 500 GeV. 
   Several comments are in order.
\begin{enumerate}
\item All normalized differential distributions, respectively the correlation coefficients are ratios. 
The P- and CP-even coefficients $C$  are, in
  the SM at NLOW and to first order in the anomalous couplings, schematically of the form
\[ C = \frac{N_0+\delta N_1+ \delta N_{\rm NP}}{\sigma_0+ \delta \sigma_1 + \delta\sigma_{\rm NP}} \, , \]
where $N_0$ $(\delta N_1)$ and $\sigma_0$ $(\delta \sigma_1)$ are the contributions  at LO QCD (NLOW). A Taylor expansion at NLO in the SM and anomalous couplings yields
\begin{equation} \label{eq:Cexp}
C  = \left(1-\frac{\delta \sigma_1}{\sigma_0}  - \frac{\delta \sigma_{\rm NP}}{\sigma_0} \right)\frac{N_0}{\sigma_0} +
\frac{\delta N_1}{\sigma_0}  + \frac{\delta N_{\rm NP}}{\sigma_0}    + \mathcal{O}(\delta^2)        \, .
\end{equation}
We use this expansion which is in the spirit of perturbation theory\footnote{ Ref.~\cite{Franzosi:2015osa} analyzed the CMDM contribution $\Hmu$
 to the helicity correlation $C(k,k)$ including the QCD corrections to this anomalous contribution. It was found that these QCD corrections
  do not improve the sensitivity to $\Hmu$.}.
\item
As expected (cf.~section~\ref{sec:LHCxsec}),  the tables~\ref{tab:Ccoefnum} - \ref{Bcoefnumlowc} show that 
  for all observables the  isospin-1 NP contribution from ${\cal L}_{\rm NP, q}$ 
  is about one order of magnitude smaller than the isospin-0
   contribution. 
\item 
Within the Standard Model $C(r,r)$ is, in the inclusive case,
 the smallest of the four P- and CP-even correlations.
This can be qualitatively understood with simple tree-level arguments.
At the LHC (8TeV) the $t$ and $\bar t$ quarks in the inclusive $\ttbar$
sample are on average only moderately relativistic.
 At or very close to threshold the $\ttbar$ system produced at LO QCD by gluon fusion
is in a $^1S_0$ state. Thus, not too far away from the threshold,
the spin configuration of the $\ttbar$ with respect to the r-axes
${\ha}_r$ and ${\hb}_r$ (which point into opposite directions) is preferentially
$\uparrow\uparrow$ or $\downarrow \downarrow$, while for relativistic
$\ttbar$ the configuration $\uparrow\downarrow$ or $\downarrow
\uparrow$ dominates. This statement applies also 
 to the  $gg\to \ttbar$ contribution to the 
  helicity correlation $C(k,k)$. The net effect for the inclusive sample is that
gluon production contributes positively both to the correlation
$C(r,r)$ and $C(k,k)$, but the correlation of the $t$ and $\bar t$ spins with
respect to the r-axes is significantly smaller than their correlation
with respect to the helicity axes ${\ha}=\hk$, $\hb= -\hk$.
The $t$, $\bar t$ produced by $q{\bar q}$ annihilation make a negative
contribution both to $C(r,r)$ and $C(k,k)$, and it turns out that the magnitude of the
contribution is in both cases approximately the same.
This is because in this case, at LO QCD,  $\ttbar$ near threshold is in a $^3S_1$
state. Thus both with respect to the r-axes and the helicity axes,
the spin configuration $\uparrow\downarrow$ or $\downarrow \uparrow$
dominates, both for non-relativistic and relativistic $t$, $\bar t$.
As a result, the contributions from $gg$ and $q{\bar q}$ initial states
almost cancel in the case of $C(r,r)$, while in the case of
$C(k,k)$ the positive contribution from the $gg$ channel is significantly
larger in magnitude than the negative one from $q{\bar q}\to \ttbar.$ \\
This line of arguments suggests that the ratio of $C(r,r)$ and
$C(k,k)$ becomes larger if the $\ttbar$ sample is restricted to events
not too far away from the $\ttbar$ threshold. This is corroborated by
the numbers of table~\ref{tab:Ccoefnumlowc}.
The fact that the correlation $C(n,n)$ is rather large in the SM can
also be understood from total angular momentum conservation. \\
The measurement of these four correlations would provide sufficient
information to simultaneously determine or constrain the
CMDM $\Hmu$ and the anomalous four-quark couplings $\hat c_{VV}$
and $\hat c_1$. \\
As an aside we remark that the NLOW and the CMDM ($\Hmu$) contribution
to the helicity correlation $C(k,k)$ at 8 TeV
and the NLOW result for 14 TeV listed in table~\ref{tab:Ccoefnum}
agree with those computed in \cite{Bernreuther:2013aga}
and in \cite{Bernreuther:2010ny}, respectively, with a different PDF
set.
   \item
 The opening angle distribution, $\sigma^{-1}d\sigma/d\cos\varphi=(1-D\cos\varphi)/2$ \cite{Bernreuther:2004jv},
  which is also a useful spin correlation observable, can  be obtained from the above diagonal correlation coefficients.
   Because the reference axes defined in table~\ref{tab:axes} form orthonormal bases, we have \cite{Bernreuther:2004jv}
  \begin{equation} \label{eq:DCrel}
    D = - \frac{1}{3}\left[C(n,n) + C(r,r) + C(k,k) \right] \, .
   \end{equation}
   This relation can be applied both to the SM and the NP contributions given in the tables below.
 The correlation coefficients $D$ computed  with \eqref{eq:DCrel} agree with those computed in  \cite{Bernreuther:2013aga} 
  and \cite{Bernreuther:2010ny} for 8 TeV and 14 TeV, respectively.
\item One may also consider 
the spin correlations $C(r^*,k)$, $C(k,r^*)$, and $C(k,k^*)$, where the $k^*$ and  $r^*$ axes 
  are defined in \eqref{eq:kstardef} and \eqref{eq:rstardef}. These observables
   are spin correlation  analogues of the charge asymmetry $A_C$. They receive contributions from 
    the entries of the $\ttbar$
    production density matrices that are odd under interchange of the $t$ and $\bar t$.  In the SM these  correlations,
     which
    receive no contributions from the $gg$ initial state and no contributions, at LO QCD,
      from $q{\bar q}$ initial states,  are expected to be small. They receive NP contributions
       proportional to  the anomalous couplings ${\hat c}_{AA}$ and ${\hat c}_2$.
\item
As long as one considers CP-invariant interactions, $C(r,k)=C(k,r)$. A non-zero difference $C(r,k)-C(k,r)$ requires
 P-even but CP-odd absorptive contributions. In the SM at NLO there are no contributions  of this type. Because our NP couplings are real and
  we do not take into account radiative corrections to the NP amplitudes, there are also no NP contributions to $C(r,k)-C(k,r)$ in our approximation.
\item 
 The sums
   \begin{equation} \label{eq:Cnrpl}
    C(n,r) + C(r,n) \, , \qquad  C(n,k)  + C(k,n) 
 \end{equation}
 probe P-odd, CP-even, naive T-odd absorptive (SM) contributions.
  The coefficients $C(n,r)$ and $C(n,k)$  receive SM contributions
   from P-odd
  absorptive parts of the mixed QCD-weak corrections. We have computed
   the  respective 1-loop absorptive contributions to the 
    $gg\to \ttbar$ and $q{\bar q}\to\ttbar$ spin density matrices. 
    The resulting effects on \eqref{eq:Cnrpl} are very small, cf. tables~\ref{tab:Ccoefnum} and~\ref{tab:Ccoefnumlowc}. 
    Because all anomalous couplings are real in the effective Lagrangian approach and because we do
    not take into account radiative corrections to the NP amplitudes, there are no NP contributions to \eqref{eq:Cnrpl}.
      \item  The SM contributions referred to in the previous item cancel in 
   \begin{equation} \label{eq:Cnrmin} 
  C(n,r) -C(r,n) \, , \qquad  C(n,k) - C(k,n)  \, .
  \end{equation}
  These differences probe P-and CP-odd (dispersive) NP contributions, that is,
   non-standard CP violation. These differences in the correlation coefficients
    can be related to CP-odd triple correlations. Defining 
  \begin{align}\label{OCPnew}
 {\cal O}^{CP}_1 =  (\hlp \times \hlm)\cdot \hk \, , 
 \qquad {\cal O}^{CP}_2 =  \text{sign}(y_p) (\hlp \times \hlm)\cdot {\hr}_p \, , 
\end{align} 
 the following relations hold:
\begin{align} \label{eq:Ctriplrel}
 C(n,r) - C(r,n) =  9 \langle{\cal O}^{CP}_1\rangle \, , \qquad
 C(n,k) - C(k,n) =  - 9 \langle {\cal O}^{CP}_2 \rangle \, .
\end{align}
 Alternatively one may consider  associated CP asymmetries, which are easier to
  measure. They are related to the expectation values of \eqref{OCPnew} by \cite{Bernreuther:2013aga,Bernreuther:1998qv}:
\begin{align} \label{eq:asyCP12}
 A_i^{CP} = \frac{N_{\ell\ell}({\cal O}^{CP}_i>0) - N_{\ell\ell}({\cal O}^{CP}_i<0)}{N_{\ell\ell}} =
 \frac{9\pi}{16} \langle  {\cal O}^{CP}_i \rangle  \, , \: \: i=1,2 \, .
\end{align}
Formulae \eqref{eq:Ctriplrel} and \eqref{eq:asyCP12} hold
if no cuts on the dileptonic final states  other than a cut on the $\ttbar$ invariant  mass are applied.
 The respective numbers in Tables ~\ref{tab:Ccoefnum} and~\ref{tab:Ccoefnumlowc} show that  $C(n,r) - C(r,n)$
 is  the most sensitive spin correlation observable to search for a top-quark chromo-electric dipole moment
  $\Hd$. The CP asymmetry $A_1^{CP}$ is less sensitive to the CP-violating anomalous couplings 
   than the left-hand sides of \eqref{eq:Ctriplrel},  but experimentally more robust.  
   The measurement of the two correlations  \eqref{eq:Ctriplrel} or the two asymmetries \eqref{eq:asyCP12}
    would allow to simultaneously determine or constrain the CEDM $\Hd$ and the anomalous CP-odd gluon-top-quark coupling
     ${\hat c}_{(--)}$. \\
     Putting the coupling ${\hat c}_{(--)}=0$, the expectation value $\langle{\cal O}^{CP}_1\rangle$ at 8 TeV computed
    with the 8 TeV value for $C(n,r) - C(r,n)$
    given in table~\ref{tab:Ccoefnum} agrees with the result of \cite{Bernreuther:2013aga}.
\item As already emphasized
  above, for $pp$ collisions a classification of observables with
  respect to CP is, 
  strictly speaking, not possible because $|pp\rangle$ is not a CP
  eigenstate. However, there will be no SM contributions  from
 $q{\bar q}$ and $gg$ initial states to the CP-odd
  observables listed in  the previous item and in \eqref{eq:diffB12}  -- assuming no
 acceptance cuts or cuts which are CP-invariant. Moreover, within our approximation we found 
  no contribution from P-violating but CP-even 
   terms induced by $gq$ production. 
\item As to the longitudinal top-polarization
 coefficients $B(k), B(r)$: The sums
 \begin{equation} \label{eq:sumB12}
  B_1(k)+ B_2(k) \, , \quad B_1(r)+ B_2(r)  \, , \quad    B_1(k^*)+ B_2(k^*) \, , \quad B_1(r^*)+ B_2(r^*)
\end{equation} 
  receive contributions from the P-odd weak-interaction corrections, that is, from  the P-odd terms of the NLOW corrections.
  However, the SM-induced $t$ and $\bar t$ polarizations are very small to tiny, as the numbers in 
   table~\ref{Bcoefnum} and~\ref{Bcoefnumlowc} show. (The SM result 
   for $B_1(k)+ B_2(k)$ at 8 TeV listed in table~\ref{Bcoefnum} agrees 
   with the result computed in \cite{Bernreuther:2013aga}.) The sums \eqref{eq:sumB12} receive also
   contributions from the P-odd NP four-quark operators.
   Thus, the measurement of the four polarization observables \eqref{eq:sumB12} would allow to simultaneously determine
    or constrain four anomalous four-quark couplings, for instance  $\hat c_{VA}$, $\hat c_{AV}$, ${\hat c}_2$, and ${\hat c}_3$, 
     assuming that input on ${\hat c}_1$ is provided from the measurement of the four CP-even spin correlations, see item 3 above. \\
   In the differences
 \begin{equation} \label{eq:diffB12}
  B_1(k)- B_2(k) \, , \quad B_1(r)- B_2(r)  \, ,\quad    B_1(k^*) - B_2(k^*) \, , \quad B_1(r^*) - B_2(r^*) \, ,
\end{equation} 
 which are P- and CP-odd and $T_N$-even, the SM contributions to $B_{1,2}$ cancel.
  Non-zero differences require, in addition to CP-violating NP contributions, also
   absorptive parts. Because our anomalous couplings are real and we do not take into account
    radiative corrections to the NP contributions, these differences remain 
    zero within our approach. In \cite{Bernreuther:2013aga} the chromo moments were
     interpreted as form factors which may have an absorptive part, and the differences of the longitudinal
     $t$ and $\bar t$ polarizations were computed in terms of ${\rm Im}\Hd$.
   \item As to the  combinations
\[B_1(n) \pm  B_2(n)  \, . \]
The sum is generated by P- and CP-even absorptive terms in the differential cross section.
 In the SM such terms arise from a number of one-loop amplitudes for $gg, q {\bar q}\to \ttbar$ 
 which involve parton, weak gauge-boson, Higgs boson, and photon exchange.
  The QCD contributions given in tables ~\ref{Bcoefnum} are in accord with 
 the 8 TeV result of \cite{Bernreuther:2013aga}. (In this work, the vector perpendicular to the scattering plane was used
  in un-normalized form and a different PDF set was employed.) 
  In addition we list in tables~\ref{Bcoefnum} and~\ref{Bcoefnumlowc} also our results for the electroweak contributions,
  which arise from the absorptive parts of the mixed QCD-weak interaction  and QCD-QED 
   contributions to $g g, q{\bar q} \to \ttbar$. They turn out to be non-negligible in comparison to the QCD contribution.
    Thus, for the inclusive $\ttbar$ sample, the transverse polarization of the
     ensemble of $t$ and $\bar t$ quarks with respect to  the reference axis  $\text{sign}(y_p){\hn}_p$
     is $P_{nt}={\overline P}_{n{\bar t}}\simeq 0.5\%$. 
  In the difference $B_1(n) -  B_2(n)$
 the SM contributions cancel. The difference
  is sensitive to P-even, but CP-odd dispersive NP contributions, which in 
 the framework of our effective Lagrangian approach are generated at Born level by the operator $\mathcal O^g_{(-+)}$ defined in \eqref{eq:Ogmp}.
 \item Finally we add a remark on the scale dependence of our SM predictions for the correlation and polarization observables.
  As the numbers given in Tables~\ref{Bcoefnum} and~\ref{Bcoefnumlowc} show, the parity-odd combinations of the $B$ coefficients are very small in the SM
   but exhibit large scale variations relative to the values for $\mu=m_t$. The reason for this is as follows. For the numerators $N$ of the $B$
   coefficients that are generated by the P-odd weak interaction contributions, one gets $N(\mu=2m_t)>N(\mu=m_t)>N(\mu=m_t/2)$, due to
    factorization-scale dependent logarithms in some of the $qg$ induced mixed QCD-weak corrections \cite{Bernreuther:2008md}, 
      while for the denominators  $D$ of the $B$ coefficients, which are taken into account at lowest order QCD, one has
      $D(\mu=2m_t)<D(\mu=m_t)<D(\mu=m_t/2)$. This produces the large scale variations in the ratios $B=N/D$. 
        The P- and CP-even correlation observables given in Tables~\ref{tab:Ccoefnum} and~\ref{tab:Ccoefnumlowc} are generated predominantly by QCD.
         Here the scale variations are moderate to small and, as is well known,  the Taylor expansion \eqref{eq:Cexp}
          reduces the scale dependence of the
         results at NLO QCD with respect to LO. The weak interaction contributions with their rather large scale dependence do not spoil
          this because they are small compared to the QCD contributions. The correlation $C(r,r)$ is special in that there are cancellations 
           among separate contributions, keeping $C(r,r)$ rather small and more sensitive to scale variations than  $C(n,n)$ and $C(k,k)$.
\end{enumerate}

The above spin-correlation and polarization observables probe all entries of the top-spin dependent parts of the $\ttbar$ production density matrices.
 In addition, the various NP contributions are disentangled to a large extent, as the above discussion and tables show.
 Moreover, a measurement of this set of spin observables -- and experimental results on $A_C$, which yields information on the coupling ${\hat c}_{AA}$ --
  would provide sufficient input for determining all anomalous couplings
  of the effective Lagrangian \eqref{eq:LNPsum}. As mentioned above, the spin correlations $C(r^*,k)$, $C(k,r^*)$, and $C(k,k^*)$ 
   are also sensitive to ${\hat c}_{AA}$.
   
   An important issue is the range of validity of the linear approximation for the NP contributions to the correlation and polarization observables.
   In analogy to the procedure we applied to  the $\ttbar$ cross section and the charge asymmetry $A_C$ in section~\ref{sec:LHCxsec},
   -- cf. Eq.  \eqref{eq:maxdiff} and Table~\ref{tab:maxdiff1}  --   
   we  estimate the maximal difference between (future) experimental measurements and SM prediction of an observable by adding
    the  moduli of the products of the respective  contributions given in Tables~\ref{tab:Ccoefnum} -- \ref{Bcoefnumlowc} 
   and the maximal values of the anomalous couplings, derived in section~\ref{suse:rappb}, for which the linear approximation can be applied.
   The results of these estimates are given, for the LHC (13 TeV) in Table~\ref{tab:linaproxCBobs}. The estimates for 8 and 14 TeV can be obtained
    in completely analogous fashion. The numbers of Table~\ref{tab:linaproxCBobs} show that the range of validity of the linear approximation
     is quite large.
 
  One may compute the above spin observables also for $\ttbar$ events with large $\ttbar$ invariant mass, for instance $\mttbar> 500$ GeV.
  While there is no computational difficulty, one should, as discussed above,  in this case abandon the  linear approximation  
  in the anomalous couplings. This leads to a proliferation  of NP contributions to each of the spin-correlation and polarization
   coefficients, which will make a comparison with (future) data rather intransparent. One may pursue a pragmatic approach, namely,
    analyze only selected observables beyond the linear approximation, in particular those which turn out to be measurable 
    with moderate or small experimental uncertainties. Such an analysis is beyond the scope of this paper and will be left to a future investigation.

\newpage
 	\begin{table}[h]
		\centering
		\caption{\label{tab:Ccoefnum}   The correlation coefficients $C$  at NLOW in the SM and the non-zero contributions
          of the dimensionless anomalous couplings defined in section~\ref{sec:Lagreff}.
       The central values correspond to the scales $\mu=m_t$. The uncertainties
        refer to  $\mu=m_t/2$ and $\mu=2m_t$.}
		\begin{tabular}{|c||c|c|c|c|c|}\hline
		\multicolumn{2}{|c|}{--}		 &  NLOW 			& $\propto \hat c_{VV}$ 			& $\propto \hat c_1$ 				& $\propto\Hmu$ \\\hline\hline	
		\multirow{3}{*}{$C(n,\,n)$} 	& 8 TeV  & $0.332^{+0.002}_{-0.002}$   & $-0.150^{+0.010}_{-0.012}$ 			& $(-1.76^{+0.15}_{-0.17})\cdot 10^{-2}$ 	& $1.831^{+0.038}_{-0.036}$ \\
						& 13 TeV & $0.326^{+0.002}_{-0.002}$ 		& $(-7.99^{+0.80}_{-0.85})\cdot 10^{-2}$ 	& $(-8.21^{+0.99}_{-1.08})\cdot 10^{-3}$ 	& $2.025^{+0.025}_{-0.024}$\\
						& 14 TeV & $0.325^{+0.002}_{-0.002}$ 		& $(-7.21^{+0.75}_{-0.81})\cdot 10^{-2}$ 	& $(-7.20^{+0.94}_{-0.99})\cdot 10^{-3}$ 	& $2.047^{+0.022}_{-0.022}$\\\hline
		\multirow{3}{*}{$C(r,\,r)$} 	& 8 TeV  & $0.055^{+0.009}_{-0.007}$ 	& $-0.934^{+0.033}_{-0.037}$ 			& $-0.122^{+0.006}_{-0.006}$ 			& $2.313^{+0.032}_{-0.031}$ \\
						& 13 TeV & $0.071^{+0.008}_{-0.006}$ 		& $-0.697^{+0.027}_{-0.027}$ 			& $(-8.46^{+0.41}_{-0.43}) \cdot 10^{-2}$ 	& $2.475^{+0.020}_{-0.019}$\\
						& 14 TeV & $0.072^{+0.008}_{-0.006}$ 		& $-0.665^{+0.025}_{-0.026}$ 			& $(-7.99^{+0.39}_{-0.40}) \cdot 10^{-2}$ 	& $2.493^{+0.019}_{-0.019}$\\\hline
		\multirow{3}{*}{$C(k,\,k)$} 	& 8 TeV  & $0.318^{+0.003}_{-0.002}$ 	& $-1.570^{+0.049}_{-0.053}$ 			& $-0.209^{+0.009}_{-0.010}$ 			& $0.964^{+0.009}_{-0.009}$ \\
						& 13 TeV & $0.331^{+0.002}_{-0.002}$ 		& $-1.218^{+0.039}_{-0.040}$ 			& $-0.151^{+0.007}_{-0.007}$ 			& $0.917^{+0.006}_{-0.006}$\\
						& 14 TeV & $0.331^{+0.002}_{-0.002}$ 		& $-1.169^{+0.037}_{-0.038}$ 			& $-0.143^{+0.006}_{-0.006}$ 			& $0.911^{+0.006}_{-0.005}$\\\hline
	\multirow{3}{*}{$C(r,\,k)+C(k,\,r)$}     & 8 TeV & $-0.226^{+0.004}_{-0.004}$ &         $-0.430^{+0.016}_{-0.018}$ 			& $(-5.42^{+0.28}_{-0.30})\cdot 10^{-2}$ 	& $0.720^{+0.008}_{-0.010}$ \\
						& 13 TeV & $-0.206^{+0.002}_{-0.002}$ 		& $-0.306^{+0.014}_{-0.014}$ 			& $(-3.58^{+0.20}_{-0.20}) \cdot 10^{-2}$ 	& $0.740^{+0.001}_{-0.002}$\\
						& 14 TeV & $-0.204^{+0.004}_{-0.002}$ 		& $-0.290^{+0.012}_{-0.012}$ 			& $(-3.34^{+0.18}_{-0.20}) \cdot 10^{-2}$ 	& $0.740^{+0.002}_{-0.004}$\\\hline
		\multicolumn{2}{|c|}{--} 					& \multicolumn{4}{|c|}{NLOW} \\\hline\hline
		\multirow{3}{*}{\scriptsize{$C(n,\,r)\!+\!C(r,\,n)$}}	& 8 TeV & \multicolumn{4}{|c|}{ $(1.03^{+0.01}_{-0.01})\cdot10^{-3}$ } \\
							& 13 TeV		& \multicolumn{4}{|c|}{ $(1.06^{+0.01}_{-0.01})\cdot 10^{-3}$ } 	\\
							& 14 TeV		& \multicolumn{4}{|c|}{ $(1.07^{+0.01}_{-0.01})\cdot 10^{-3}$ } \\\hline
		\multirow{3}{*}{\scriptsize{$C(n,\,k)\!+\!C(k,\,n)$}}	& 8 TeV & \multicolumn{4}{|c|}{ $(2.32^{+0.01}_{-0.01})\cdot 10^{-3}$ } \\
							& 13 TeV		& \multicolumn{4}{|c|}{ $(2.15^{+0.04}_{-.007})\cdot 10^{-3}$ } 	\\
							& 14 TeV		& \multicolumn{4}{|c|}{ $(2.12^{+0.03}_{-0.08})\cdot 10^{-3}$ } 	\\\hline\hline
		\multicolumn{2}{|c|}{--} 					& \multicolumn{2}{|c|}{$\propto \hat c_{(--)}$} 	& \multicolumn{2}{|c|}{$\propto \Hd$} \\\hline\hline
		\multirow{3}{*}{\scriptsize{$C(n,\,r)\!-\!C(r,\,n)$}}	& 8 TeV & \multicolumn{2}{|c|}{ $-1.321^{+0.021}_{-0.020}$ } 	& \multicolumn{2}{|c|}{ $-3.705^{+0.084}_{-0.087}$ } \\
							& 13 TeV		& \multicolumn{2}{|c|}{ $-1.226^{+0.007}_{-0.004}$ } 	& \multicolumn{2}{|c|}{ $-4.143^{+0.053}_{-0.056}$ } \\
							& 14 TeV		& \multicolumn{2}{|c|}{ $-1.217^{+0.005}_{-0.002}$ } 	& \multicolumn{2}{|c|}{ $-4.195^{+0.052}_{-0.052}$ } \\\hline
		\multirow{3}{*}{\scriptsize{$C(n,\,k)\!-\!C(k,\,n)$}}	& 8 TeV & \multicolumn{2}{|c|}{ $-1.665^{+0.073}_{-0.058}$ } 	& \multicolumn{2}{|c|}{ $-0.721^{+0.021}_{-0.019}$ } \\
							& 13 TeV		& \multicolumn{2}{|c|}{ $-2.157^{+0.108}_{-0.087}$ } 	& \multicolumn{2}{|c|}{ $-0.800^{+0.006}_{-0.003}$ } \\
							& 14 TeV		& \multicolumn{2}{|c|}{ $-2.236^{+0.112}_{-0.091}$ } 	& \multicolumn{2}{|c|}{ $-0.805^{+0.004}_{-0.001}$ } \\\hline
		\end{tabular}
	\end{table}

\newpage
      
	\begin{table}[h]
		\centering
		\caption{\label{Bcoefnum} The correlation coefficients $B$ at NLOW in the SM and the non-zero contributions
          of the dimensionless anomalous couplings defined in section~\ref{sec:Lagreff}.
      The central values correspond to the scales $\mu=m_t$. The uncertainties
        refer to  $\mu=m_t/2$ and $\mu=2m_t$.}
		\begin{tabular}{|c||c|c|c|c|c|c|c|}\hline
		\multicolumn{2}{|c|}{--} 			& \multicolumn{2}{|c|}{NLOW} 			& \multicolumn{2}{|c|}{$\propto \hat c_{VA}$} 		& \multicolumn{2}{|c|}{$\propto \hat c_{3}$}  \\\hline\hline
		\multirow{3}{*}{$B_1(r) + B_2(r)$} & 8 TeV 	& \multicolumn{2}{|c|}{$(2.9^{+1.9}_{-1.4})\cdot 10^{-3}$} 	& \multicolumn{2}{|c|}{$0.298^{+0.014}_{-0.014}$} 	& \multicolumn{2}{|c|}{$(3.89^{+0.24}_{-0.23})\cdot 10^{-2}$} \\
						    & 13 TeV	& \multicolumn{2}{|c|}{$(3.2^{+2.3}_{-1.7})\cdot 10^{-3}$} 	& \multicolumn{2}{|c|}{$0.210^{+0.009}_{-0.009}$} 	& \multicolumn{2}{|c|}{$(2.55^{+0.14}_{-0.14})\cdot 10^{-2}$} \\
						    & 14 TeV	& \multicolumn{2}{|c|}{$(3.3^{+2.3}_{-1.8})\cdot 10^{-3}$} 	& \multicolumn{2}{|c|}{$0.199^{+0.008}_{-0.008}$} 	& \multicolumn{2}{|c|}{$(2.39^{+0.13}_{-0.13})\cdot 10^{-2}$} \\\hline
		\multirow{3}{*}{$B_1(k) + B_2(k)$} & 8 TeV 	& \multicolumn{2}{|c|}{$(6.1^{+2.3}_{-1.7}) \cdot 10^{-3}  $} 	& \multicolumn{2}{|c|}{$2.100^{+0.081}_{-0.078}$} 	& \multicolumn{2}{|c|}{$0.283^{+0.015}_{-0.014}$} \\
						    & 13 TeV	& \multicolumn{2}{|c|}{$(8.0^{+3.4}_{-2.4})\cdot 10^{-3}$} 	& \multicolumn{2}{|c|}{$1.607^{+0.051}_{-0.052}$} 	& \multicolumn{2}{|c|}{$0.201^{+0.009}_{-0.009}$} \\
						    & 14 TeV	& \multicolumn{2}{|c|}{$(8.3^{+3.5}_{-3.5})\cdot 10^{-3}$} 	& \multicolumn{2}{|c|}{$1.542^{+0.047}_{-0.047}$} 	& \multicolumn{2}{|c|}{$0.191^{+0.008}_{-0.008}$} \\\hline\hline
		\multicolumn{2}{|c|}{--} 			& \multicolumn{2}{|c|}{NLOW }			& \multicolumn{2}{|c|}{$\propto \hat c_{AV}$} 		& \multicolumn{2}{|c|}{$\propto \hat c_{1} - \hat c_{2} + \hat c_{3}$}   \\\hline\hline
		\multirow{3}{*}{$B_1(r^*) + B_2(r^*)$}	& 8 TeV & \multicolumn{2}{|c|}{$< 10^{-3}$} 	& \multicolumn{2}{|c|}{$1.166^{+0.063}_{-0.060}$} 	& \multicolumn{2}{|c|}{$0.244^{+0.014}_{-0.014}$}   \\
							& 13 TeV& \multicolumn{2}{|c|}{$< 10^{-3} $} 	& \multicolumn{2}{|c|}{$0.792^{+0.040}_{-0.039}$} 	& \multicolumn{2}{|c|}{$0.164^{+0.009}_{-0.009}$} \\
							& 14 TeV& \multicolumn{2}{|c|}{$< 10^{-3} $} 	& \multicolumn{2}{|c|}{$0.746^{+0.037}_{-0.036}$} 	& \multicolumn{2}{|c|}{$0.154^{+0.008}_{-0.008}$} \\\hline
		\multirow{3}{*}{$B_1(k^*) + B_2(k^*)$}	& 8 TeV & \multicolumn{2}{|c|}{$ < 10^{-3} $} 	& \multicolumn{2}{|c|}{$1.204^{+0.054}_{-0.052}$} 	& \multicolumn{2}{|c|}{$0.251^{+0.013}_{-0.012}$}  \\
							& 13 TeV& \multicolumn{2}{|c|}{$< 10^{-3}   $} 	& \multicolumn{2}{|c|}{$0.883^{+0.034}_{-0.034}$} 	& \multicolumn{2}{|c|}{$0.183^{+0.008}_{-0.008}$} \\
							& 14 TeV& \multicolumn{2}{|c|}{$ < 10^{-3}  $} 	& \multicolumn{2}{|c|}{$0.842^{+0.032}_{-0.032}$} 	& \multicolumn{2}{|c|}{$0.174^{+0.007}_{-0.007}$} \\\hline
		\multicolumn{2}{|c|}{--} & \multicolumn{3}{|c|}{NLO QCD}& \multicolumn{3}{|c|}{EW} \\\hline\hline
		\multirow{3}{*}{$B_1(n) + B_2(n)$}	& 8 TeV &  \multicolumn{3}{|c|}{$(6.93^{+0.84}_{-0.72})\cdot 10^{-3}$} &  \multicolumn{3}{|c|}{$(3.44^{+0.09}_{-0.07})\cdot 10^{-3}$}\\
							& 13 TeV&  \multicolumn{3}{|c|}{$(7.45^{+0.90}_{-0.72})\cdot 10^{-3}$} &  \multicolumn{3}{|c|}{$(3.89^{+0.06}_{-0.06})\cdot 10^{-3}$}\\
							& 14 TeV&  \multicolumn{3}{|c|}{$(7.52^{+0.90}_{-0.73})\cdot 10^{-3}$} &  \multicolumn{3}{|c|}{$(3.94^{+0.06}_{-0.05})\cdot 10^{-3}$}\\\hline
		\multicolumn{2}{|c|}{--} & \multicolumn{6}{|c|}{$\propto \hat c_{(-+)}$} \\\hline\hline
		\multirow{3}{*}{$B_1(n) - B_2(n)$}	& 8 TeV &  \multicolumn{6}{|c|}{$4.148^{+0.115}_{-0.128}$} \\
							& 13 TeV&  \multicolumn{6}{|c|}{$4.876^{+0.108}_{-0.124}$} \\
							& 14 TeV&  \multicolumn{6}{|c|}{$4.975^{+0.109}_{-0.124}$} \\\hline
		\end{tabular}
		\end{table}

\newpage
   
	\begin{table}[h]
		\centering
		\caption{\label{tab:Ccoefnumlowc}   The correlation coefficients $C$  at NLOW in the SM and the non-zero contributions
          of the dimensionless anomalous couplings defined in section~\ref{sec:Lagreff} for  $\mttbar \leq$ 500 GeV. }
		\begin{tabular}{|c||c|c|c|c|c|}\hline
		\multicolumn{2}{|c|}{--}		 &  NLOW 			& $\propto \hat c_{VV}$ 			& $\propto \hat c_1$ 				& $\propto\Hmu$ \\\hline\hline	
		\multirow{3}{*}{$C(n,\,n)$} 	& 8 TeV  & $0.396^{+0.002}_{-0.002}$ 		& $-0.317^{+0.012}_{-0.013}$		 	& $(-3.89^{+0.20}_{-0.22})\cdot 10^{-2}$ 	& $1.864^{+0.035}_{-0.034}$ \\
						& 13 TeV & $0.403^{+0.002}_{-0.002}$ 		& $-0.230^{+0.009}_{-0.010}$		 	& $(-2.59^{+0.14}_{-0.14})\cdot 10^{-2}$ 	& $2.041^{+0.024}_{-0.023}$\\
						& 14 TeV & $0.403^{+0.002}_{-0.002}$ 		& $-0.219^{+0.009}_{-0.009}$ 			& $(-2.43^{+0.13}_{-0.13})\cdot 10^{-2}$ 	& $2.062^{+0.023}_{-0.022}$\\\hline
		\multirow{3}{*}{$C(r,\,r)$} 	& 8 TeV  & $0.137^{+0.007}_{-0.005}$ 		& $-0.821^{+0.030}_{-0.032}$ 			& $-0.101^{+0.005}_{-0.005}$ 			& $2.360^{+0.031}_{-0.028}$ \\
						& 13 TeV & $0.168^{+0.005}_{-0.003}$ 		& $-0.599^{+0.024}_{-0.025}$ 			& $(-6.78^{+0.35}_{-0.36}) \cdot 10^{-2}$ 	& $2.510^{+0.020}_{-0.020}$\\
						& 14 TeV & $0.171^{+0.005}_{-0.003}$ 		& $-0.571^{+0.022}_{-0.023}$ 			& $(-6.36^{+0.33}_{-0.33}) \cdot 10^{-2}$ 	& $2.527^{+0.019}_{-0.018}$\\\hline
		\multirow{3}{*}{$C(k,\,k)$} 	& 8 TeV  & $0.413^{+0.003}_{-0.003}$ 		& $-1.060^{+0.038}_{-0.040}$ 			& $-0.131^{+0.006}_{-0.007}$ 			& $0.928^{+0.010}_{-0.013}$ \\
						& 13 TeV & $0.446^{+0.007}_{-0.008}$ 		& $-0.777^{+0.030}_{-0.031}$ 			& $(-8.80^{+0.44}_{-0.46})\cdot 10^{-2}$ 	& $0.863^{+0.009}_{-0.010}$\\
						& 14 TeV & $0.448^{+0.009}_{-0.008}$ 		& $-0.741^{+0.028}_{-0.029}$ 			& $(-8.28^{+0.41}_{-0.43})\cdot 10^{-2}$ 	& $0.855^{+0.008}_{-0.010}$\\\hline
	\multirow{3}{*}{$C(r,\,k)+C(k,\,r)$}    & 8 TeV & $-0.230^{+0.004}_{-0.004}$ 	       & $-0.430^{+0.016}_{-0.016}$		 	& $(-5.30^{+0.26}_{-0.28})\cdot 10^{-2}$ 	& $0.980^{+0.014}_{-0.014}$ \\
						& 13 TeV & $-0.212^{+0.010}_{-0.002}$ 		& $-0.316^{+0.012}_{-0.012}$ 			& $(-3.56^{+0.18}_{-0.18}) \cdot 10^{-2}$ 	& $1.054^{+0.010}_{-0.010}$\\
						& 14 TeV & $-0.210^{+0.002}_{-0.002}$ 		& $-0.300^{+0.012}_{-0.012}$ 			& $(-3.36^{+0.16}_{-0.16}) \cdot 10^{-2}$ 	& $1.062^{+0.010}_{-0.010}$\\\hline
		\multicolumn{2}{|c|}{--} 					& \multicolumn{4}{|c|}{NLOW} \\\hline\hline
		\multirow{3}{*}{\scriptsize{$C(n,\,r)\!+\!C(r,\,n)$}}	& 8 TeV & \multicolumn{4}{|c|}{ $< 10^{-3}$ } \\
							& 13 TeV		& \multicolumn{4}{|c|}{ $ < 10^{-3}$ } 	\\
							& 14 TeV		& \multicolumn{4}{|c|}{ $ < 10^{-3} $ } \\\hline
		\multirow{3}{*}{\scriptsize{$C(n,\,k)\!+\!C(k,\,n)$}}	& 8 TeV & \multicolumn{4}{|c|}{ $(2.64^{+0.04}_{-0.05})\cdot 10^{-3}$ } \\
							& 13 TeV		& \multicolumn{4}{|c|}{ $(2.87^{+0.04}_{-0.03})\cdot 10^{-3}$ } 	\\
							& 14 TeV		& \multicolumn{4}{|c|}{ $(2.91^{+0.02}_{-0.04})\cdot 10^{-3}$ } 	\\\hline\hline
		\multicolumn{2}{|c|}{--} 					& \multicolumn{2}{|c|}{$\propto \hat c_{(--)}$} 	& \multicolumn{2}{|c|}{$\propto \Hd$} \\\hline\hline
		\multirow{3}{*}{\scriptsize{$C(n,\,r)\!-\!C(r,\,n)$}}	& 8 TeV & \multicolumn{2}{|c|}{ $-0.667^{+0.035}_{-0.035}$ } 	& \multicolumn{2}{|c|}{ $-3.746^{+0.086}_{-0.091}$ } \\
							& 13 TeV		& \multicolumn{2}{|c|}{ $-0.481^{+0.023}_{-0.023}$ } 	& \multicolumn{2}{|c|}{ $-4.208^{+0.061}_{-0.062}$ } \\
							& 14 TeV		& \multicolumn{2}{|c|}{ $-0.459^{+0.021}_{-0.022}$ } 	& \multicolumn{2}{|c|}{ $-4.261^{+0.057}_{-0.059}$ } \\\hline
		\multirow{3}{*}{\scriptsize{$C(n,\,k)\!-\!C(k,\,n)$}}	& 8 TeV & \multicolumn{2}{|c|}{ $0.719^{+0.036}_{-0.035}$ } 	& \multicolumn{2}{|c|}{ $-1.098^{+0.032}_{-0.034}$ } \\
							& 13 TeV		& \multicolumn{2}{|c|}{ $0.908^{+0.022}_{-0.022}$ } 	& \multicolumn{2}{|c|}{ $-1.274^{+0.023}_{-0.023}$ } \\
							& 14 TeV		& \multicolumn{2}{|c|}{ $0.929^{+0.020}_{-0.020}$ } 	& \multicolumn{2}{|c|}{ $-1.294^{+0.021}_{-0.022}$ } \\\hline
		\end{tabular}
	\end{table}
      
   \newpage   
       
	\begin{table}[h]
		\centering
		\caption{\label{Bcoefnumlowc} The correlation coefficients $B$ at NLOW in the SM and the non-zero contributions
          of the dimensionless anomalous couplings defined in section~\ref{sec:Lagreff}  for $\mttbar \leq$ 500 GeV.}
		\begin{tabular}{|c||c|c|c|c|c|c|}\hline
		\multicolumn{2}{|c|}{--} 			& NLOW 			& \multicolumn{2}{|c|}{$\propto \hat c_{VA}$} 		& \multicolumn{2}{|c|}{$\propto \hat c_{3}$}  \\\hline\hline
		\multirow{3}{*}{$B_1(r) + B_2(r)$} & 8 TeV 	& $(2.5^{+2.3}_{-2.1})\cdot 10^{-3}$ 	& \multicolumn{2}{|c|}{$0.196^{+0.010}_{-0.010}$} 	& \multicolumn{2}{|c|}{$(2.43^{+0.16}_{-0.16})\cdot 10^{-2}$} \\
						    & 13 TeV	& $(2.9^{+2.8}_{-1.9})\cdot 10^{-3}$ 	& \multicolumn{2}{|c|}{$0.134^{+0.006}_{-0.006}$} 	& \multicolumn{2}{|c|}{$(1.52^{+0.09}_{-0.09})\cdot 10^{-2}$} \\
						    & 14 TeV	& $(2.9^{+2.9}_{-2.1})\cdot 10^{-3}$ 	& \multicolumn{2}{|c|}{$0.127^{+0.006}_{-0.006}$} 	& \multicolumn{2}{|c|}{$(1.42^{+0.08}_{-0.08})\cdot 10^{-2}$} \\\hline
		\multirow{3}{*}{$B_1(k) + B_2(k)$} & 8 TeV 	& $(2.6^{+2.0}_{-1.5})\cdot 10^{-3}$ 	& \multicolumn{2}{|c|}{$0.979^{+0.050}_{-0.049}$} 	& \multicolumn{2}{|c|}{$0.121^{+0.008}_{-0.008}$} \\
						    & 13 TeV	& $(3.5^{+2.7}_{-2.0})\cdot 10^{-3}$ 	& \multicolumn{2}{|c|}{$0.669^{+0.032}_{-0.031}$} 	& \multicolumn{2}{|c|}{$(7.60^{+0.45}_{-0.45})\cdot 10^{-2}$} \\
						    & 14 TeV	& $(3.6^{+2.8}_{-2.7})\cdot 10^{-3}$ 	& \multicolumn{2}{|c|}{$0.632^{+0.029}_{-0.029}$} 	& \multicolumn{2}{|c|}{$(7.09^{+0.42}_{-0.41})\cdot 10^{-2}$} \\\hline\hline
		\multicolumn{2}{|c|}{--} 			& NLOW 			& \multicolumn{2}{|c|}{$\propto \hat c_{AV}$} 		& \multicolumn{2}{|c|}{$\propto \hat c_{1} - \hat c_{2} + \hat c_{3}$}   \\\hline\hline
		\multirow{3}{*}{$B_1(r^*) + B_2(r^*)$}	& 8 TeV & $<10^{-3}$ 	& \multicolumn{2}{|c|}{$0.988^{+0.055}_{-0.053}$} 	& \multicolumn{2}{|c|}{$0.207^{+0.012}_{-0.012}$}   \\
							& 13 TeV& $<10^{-3}$ 	& \multicolumn{2}{|c|}{$0.659^{+0.034}_{-0.033}$} 	& \multicolumn{2}{|c|}{$0.135^{+0.008}_{-0.007}$} \\
							& 14 TeV& $<10^{-3}$ 	& \multicolumn{2}{|c|}{$0.620^{+0.031}_{-0.031}$} 	& \multicolumn{2}{|c|}{$0.126^{+0.007}_{-0.007}$} \\\hline
		\multirow{3}{*}{$B_1(k^*) + B_2(k^*)$}	& 8 TeV & $<10^{-3}$ 	& \multicolumn{2}{|c|}{$0.765^{+0.042}_{-0.041}$} 	& \multicolumn{2}{|c|}{$0.160^{+0.009}_{-0.010}$}  \\
							& 13 TeV& $<10^{-3}$ 	& \multicolumn{2}{|c|}{$0.511^{+0.026}_{-0.025}$} 	& \multicolumn{2}{|c|}{$0.105^{+0.006}_{-0.006}$} \\
							& 14 TeV& $<10^{-3}$ 	& \multicolumn{2}{|c|}{$0.481^{+0.024}_{-0.024}$} 	& \multicolumn{2}{|c|}{$(9.83^{+0.05}_{-0.05})\cdot 10^{-2}$} \\\hline
		\multicolumn{2}{|c|}{--} & \multicolumn{2}{|c|}{NLO QCD}& \multicolumn{3}{|c|}{EW} \\\hline\hline
		\multirow{3}{*}{$B_1(n) + B_2(n)$}	& 8 TeV &  \multicolumn{2}{|c|}{$(4.12^{+0.44}_{-0.36})\cdot 10^{-3}$} &  \multicolumn{3}{|c|}{$(3.14^{+0.09}_{-0.05})\cdot 10^{-3}$}\\
							& 13 TeV&  \multicolumn{2}{|c|}{$(4.16^{+0.45}_{-0.37})\cdot 10^{-3}$} &  \multicolumn{3}{|c|}{$(3.50^{+0.07}_{-0.03})\cdot 10^{-3}$}\\
							& 14 TeV&  \multicolumn{2}{|c|}{$(4.16^{+0.45}_{-0.37})\cdot 10^{-3}$} &  \multicolumn{3}{|c|}{$(3.55^{+0.05}_{-0.03})\cdot 10^{-3}$}\\\hline
		\multicolumn{2}{|c|}{--} & \multicolumn{5}{|c|}{$\propto \hat c_{(-+)}$} \\\hline\hline
		\multirow{3}{*}{$B_1(n) - B_2(n)$}	& 8 TeV &  \multicolumn{5}{|c|}{$2.499^{+0.047}_{-0.051}$} \\
							& 13 TeV&  \multicolumn{5}{|c|}{$2.754^{+0.036}_{-0.038}$} \\
							& 14 TeV&  \multicolumn{5}{|c|}{$2.785^{+0.034}_{-0.036}$} \\\hline
		\end{tabular}
		\end{table}

 \newpage

  \begin{table}[h]
	\caption{Moduli of maximal differences between future experimental results and SM predictions, for 
	 the correlation and polarization observables at the LHC (13 TeV), for which the 
	linear approximation can be applied.}
	\centering
  	\begin{tabular}{|c||c|c|}\hline
  	Observable & no cut & $m_{t\bar t} < 500 \text{GeV}$ \\\hline\hline
  		$\delta C(n, \, n)$ & 0.31 & 0.33 \\\hline
  		$\delta C(r, \, r)$ & 0.44 & 0.44\\\hline
  		$\delta C(k, \, k)$ & 0.27 & 0.22\\\hline
  		$\delta [C(r, \, k)+C(k,\,r)]$ & 0.15 & 0.19  \\\hline
  		$\delta [C(n, \, r)-C(r,\,n)]$ & 0.74 & 0.68\\\hline
  		$\delta [C(n, \, k)-C(k,\,n)]$ & 0.34 & 0.28 \\\hline
  		$\delta [B_1(r)+B_2(r)]$ & 2.4$\times 10^{-2}$ &  1.5$\times 10^{-2}$  \\\hline
  		$\delta [B_1(k)+B_2(k)]$ & 0.18& 7.5$\times 10^{-2}$\\\hline
  		$\delta [B_1(r^*)+B_2(r^*)]$ & 0.13 & 0.11 \\\hline
  		$\delta [B_1(k^*)+B_2(k^*)]$ & 0.14 & 8.3$\times 10^{-2}$\\\hline
  		$\delta [B_1(n)-B_2(n)]$ & 0.49 & 0.28\\\hline
  	\end{tabular}
  	\label{tab:linaproxCBobs}
  \end{table}
 
\section{Summary and conclusions}
 \label{sec:sumconc}

 We have defined a set of correlation and polarization observables with which hadronic $\ttbar$ production dynamics can be probed in detail, that is, all
  entries of the top-spin dependent parts of the  hadronic $\ttbar$ production density matrices, in dileptonic and semileptonic $\ttbar$ events at the LHC.
  Our observables allow, in particular,  to disentangle contributions from interactions which conserve/violate parity and/or CP. 
   We have computed these observables within the Standard Model at NLO QCD including the  mixed QCD-(electro)weak contributions. In addition we have 
    analyzed possible new physics contributions in terms of an effective Lagrangian that
     contains all gauge-invariant effective interactions of mass dimension $\leq 6$
     being relevant for the parton processes at hand. Existing constraints 
      on anomalous couplings which describe the strength of these effective NP interactions justify that 
      only NP terms linear in the anomalous couplings are taken into account in the computation of  inclusive (spin) observables
       for the LHC.  Irrespective of these constraints we have estimated the range of validity of the linear approximation
        for the observables considered in this paper. 
         Some of our polarization and spin correlation observables allow for direct searches of non-standard P- and CP violation in
         $\ttbar$ events  and will provide, once measurements will have been made, 
          direct information on respective interactions which may affect hadronic $\ttbar$ production. The measurement of 
           the complete set of our inclusive spin observables, including $A_C$,
            would allow to determine, respectively constrain all anomalous couplings contained in ${\cal L}_{\rm NP}$.
          Instead of $A_C$, one may  consider the observables  $C(r^*,k)$, $C(k,r^*)$, and $C(k,k^*)$ 
           which are the spin correlation analogues of the charge asymmetry.
           
               Eventually, it will also be of interest to compute and  measure the above $\ttbar$ spin correlations and $t$ and $\bar t$ polarizations 
          in the kinematic regime where the energy-momentum transfer to the $\ttbar$ system is large,
           for instance for $\ttbar$ events  with $\mttbar> 500$ GeV.
            In this regime the contributions of the effective
           NP interactions grow in general, due to their non-renormalizable nature. 
            In this kinematic regime one should  abandon the linear approximation, but then each observable
             receives a large number of contributions from the 
             anomalous interactions contained in ${\cal L}_{\rm NP}$. Such an analysis should be done at some point for selected 
              observables, once experimental feedback on measurement precisions will be available.

\appendix
\section{NP contributions to the $\ttbar$ spin density matrices}
\label{sec:AppA}
 The $\ttbar$ production spin density matrices to the parton reactions
      $gg, q{\bar q} \to \ttbar (g)$ and $g q ({\bar q})\to \ttbar q ({\bar q})$ 
       were computed at NLO QCD (dispersive contributions) in \cite{Bernreuther:2001rq,Bernreuther:2004jv}, the contributions of the NLO QCD absorptive
        parts were determined in \cite{Bernreuther:1995cx,Dharmaratna:1996xd} and the contributions of the mixed QCD-weak corrections (dispersive contributions)
         in \cite{Bernreuther:2005is,Bernreuther:2006vg,Bernreuther:2008md}. The spin-independent term of the mixed QCD-weak corrections to
          $q{\bar q}, gg \to \ttbar$, i.e., the respective contributions to the coefficient functions $A^I$ of eq.~\eqref{Rot}, were computed also in  \cite{Kuhn:2005it,Kuhn:2006vh} and, without the infrared-divergent
           box contributions to $q{\bar q}\to \ttbar$ and the corresponding real gluon radiation, 
          in \cite{Beenakker:1993yr}, 
          and these results agree with ours. 
           In addition, we have computed the contributions of the absorptive parts of
          the mixed QCD electroweak corrections. Here we do not list the latter contributions, for the sake of brevity and because
           they induce only very small effects (see section~\ref{sec:corrllnew}). Semileptonic and non-leptonic decays of polarized
            top-quarks are incorporated at NLO QCD \cite{Brandenburg:2002xr}.

 Here we list the contributions to the $\ttbar$ spin density matrices \eqref{Rot} 
     in the decomposition \eqref{eq:newB}, \eqref{eq:newC}
  of the NP interactions described by the effective Lagrangian \eqref{eq:LNPsum}.
  As emphasized above, we take into account only the interferences with the 
   tree-level QCD amplitudes of $gg, q{\bar q} \to \ttbar$, i.e., only terms
    linear in the anomalous couplings. We recall that all anomalous couplings are real dimensionless
     parameters. \\
      In the following, $s=(p_1+p_2)^2$, where $p_{1,2}$ are the 4-momenta of the initial partons. Furthermore
       we define (cf. \eqref{orhtoset})
       \begin{equation}\label{eq:defyzbet}
        z= \frac{2 m_t}{\sqrt s} \, , \qquad \beta=\sqrt{1-z^2} \, , \quad y = \mathbf{\hat p}\cdot \mathbf{\hat k} \, .
       \end{equation}

   \subsection{NP contributions to $g g\to \ttbar$}    
     For the non-vanishing NP contributions to the coefficients of the $g g\to \ttbar$ spin density matrix  defined in \eqref{eq:newB}, \eqref{eq:newC}
      we get in the linear approximation, dropping   the superscript $I=g g$:  
      
  	\begin{align}
	A = &\; \frac{N_c^2(1+\beta^2 y^2)-2}{1-\beta^2 y^2} \; \Hmu,\\
	b^{+}_n =& -b^{-}_n = 2 N_c^2 \,\frac{y\beta^2}{z}\,\frac{\sqrt{1-y^2}}{\bigl(1-\beta^2 y^2\bigr)}\; {\hat c}_{(-+)}, \\
	c_{rr} = & \; \frac{
	 N_c^2 \left[-1+\beta^4 y^2 \bigl(1-y^2\bigr)^2 + \beta^2 \bigl(1-y^2+y^4\bigr)\right] + 2 - 2\beta^2\bigl(1-y^2+y^4\bigr)}{\bigl(1-\beta^2 y^2\bigr)^2} \; \Hmu,\\
	  c_{kk} =&\; N_c^2 \frac{-1-\beta^2\bigl(-2+y^4\bigr) + \beta^6 y^2 \bigl(-1-y^2+y^4\bigr) + 
	  \beta^4 \bigl(-1+y^2+2 y^4 - y^6\bigr)}{z^2\bigl(1-\beta^2 y^2\bigr)^2}\; \Hmu \nonumber \\
	 & \; +  \frac{2}{z^2}\,\frac{1+\beta^4\bigl(1+y^2 - y^4\bigr) + \beta^2 \bigl(-2-y^2 + y^4\bigr)}{\bigl(1-\beta^2 y^2\bigr)^2}\; \Hmu,
	 \end{align}
	 \begin{align}
	c_{nn} =&\; \frac{2-N_c^2}{1-\beta^2 y^2} \; \Hmu,\\ 
	c_{rk}=&\; N_c^2 \frac{y\sqrt{1-y^2}}{2z}\,\frac{-2 \beta^2 y^2 + \beta^6 y^2 \bigl(-1 + y^2\bigr) + \beta^4 \bigl(1+3 y^2 - 2 y^4\bigr)}{\bigl(1-\beta^2 y^2\bigr)^2}\;\Hmu \nonumber \\
	&\; -\frac{y \beta^2\sqrt{1-y^2}}{z}\,\frac{1+ y^2\bigl(-2+\beta^2\bigr)}{\bigl(1-\beta^2 y^2\bigr)^2}\; \Hmu \, ,
	\end{align}
	\begin{align}
	c_r &= -\frac{N_c^2 y\beta}{z^3}\,\frac{\sqrt{1-y^2}\Bigl(1-3 z^4 + \beta^2 \bigl(z^2-z^4 y^2\bigr) + \beta^6 \bigl(y^2+y^4\bigr) - \beta^4\bigl(1 + y^2 + y^2 z^2 + y^4\bigr)\Bigr)}{\bigl(1-\beta^2 y^2\bigr)^2}
	   \;{\hat c}_{(--)} \nonumber \\
	&\:\:\:\:\:-\frac{N_c^2 y\beta\sqrt{1-y^2}}{2z}\,\frac{2+\beta^2\bigl(-3+y^2\bigr) + \beta^4 y^2 \bigl(-1+y^2\bigr)}{\bigl(1-y^2\beta^2\bigr)^2}\; \Hd \nonumber \\
	 & \:\:\:\:\:-2\,\frac{y\beta}{z}\,\frac{\sqrt{1-y^2}\Bigl(1+\beta^2\bigl(y^2-2\bigr)\Bigr)}{\bigl(1-\beta^2 y^2\bigr)^2}\;{\hat c}_{(--)} 
	\; + \; \frac{y\beta\sqrt{1-y^2}}{z}\,\frac{1+\beta^2\bigl(-2+y^2\bigr)}{\bigl(1-\beta^2 y^2\bigr)^2}\; \Hd,
	\end{align}
	\begin{align}
	c_k &= -\frac{N_c^2 \beta}{z^2\bigl(1-\beta^2 y^2\bigr)^2}\,
	\Bigl[ -1+y^2+z^4\bigl(2-3 y^2\bigr) + \beta^6 y^4 \bigl(-1+y^2\bigr) - \beta^2 \bigl(-1 + y^2 -2 y^2 z^4 + y^4 z^4\bigr) \nonumber \\
	&\:\:\:\:\:+ \beta^4 \bigl(y^4-y^6\bigr)\Bigr]\;{\hat c}_{(--)} \: - \: \frac{N_c^2 \beta}{2}\,\frac{-3+2 y^2 + \beta^2 \bigl(2-3 y^2 + y^4\bigr) + \beta^4 y^2 
	 \bigl(2- 2 y^2 + y^4\bigr)}{\bigl(1-\beta^2 y^2\bigr)^2}\; \Hd \nonumber \\
	 &  \:\:\:\:\: + \; 2 \beta\bigl(1-y^2\bigr)\, \frac{1+ \beta^2\bigl(-2+y^2\bigr)}{\bigl(1 -\beta^2 y^2\bigr)^2}\;{\hat c}_{(--)} 
	  \:+\: \beta\,\frac{-3 + y^2 + \beta^2 \bigl(2 - y^2 + y^4\bigr)}{\bigl(1-\beta^2 y^2\bigr)^2}\;\Hd \, .
	\end{align}     

   \subsection{NP contributions to $q{\bar q} \to \ttbar$}   
 
     For the non-vanishing NP contributions to the coefficients of the $q{\bar q} \to \ttbar$ 
     spin density matrix  defined in \eqref{eq:newB}, \eqref{eq:newC}
      we get in the linear approximation, dropping   the superscript $I=q{\bar q}$:
       \begin{eqnarray}
	 A = & \; \displaystyle{\frac{2 + \beta^2 \bigl(-1 + y^2\bigr)}{4 z^2} \; \bigl[{\hat c}_{VV} \pm \frac{{\hat c}_1}{2} \bigr] \:+\: \frac{y \beta}{2 z^2} \; \bigl[{\hat c}_{AA}  \pm \frac{{\hat c}_2}{2} \bigr] \: + \: \frac{1}{2}\;\Hmu,} \\
     b^{+}_r = & \, b^{-}_r \,=\; \displaystyle{\frac{y\beta\sqrt{1-y^2}}{4 z} \; [{\hat c}_{VA} \pm \frac{{\hat c}_3}{2}]   \:+ \: \frac{\sqrt{1-y^2}}{2z} \;  [{\hat c}_{AV}\pm \frac{{\hat c}_1}{2} \mp \frac{{\hat c}_2}{2} \pm \frac{{\hat c}_3}{2}] } ,\\
  b^{+}_k = &\, b^{-}_k \,=\; \displaystyle{\frac{\beta (1+y^2)}{4 z^2}\; [{\hat c}_{VA} \pm \frac{{\hat c}_3}{2}]   \:+\: \frac{y}{2 z^2} [{\hat c}_{AV}\pm \frac{{\hat c}_1}{2} \mp \frac{{\hat c}_2}{2} \pm \frac{{\hat c}_3}{2}]\; }
	   \, , \\
     c_{rr} = &\; \displaystyle{-\frac{(2-\beta^2)(1-y^2)}{4 z^2} \; [{\hat c}_{VV} \pm \frac{{\hat c}_1}{2}] \: + \: \frac{1}{2}(1-y^2)\; \Hmu } \, , \\
	c_{kk} = &\; \displaystyle{\frac{(2 y^2 + \beta^2 (1-y^2))}{4 z^2} \; [{\hat c}_{VV} \pm \frac{{\hat c}_1}{2}] 
	 \:+\: \frac{y\beta}{2 z^2}\; [{\hat c}_{AA}  \pm \frac{{\hat c}_2}{2} ]  \: + \: \frac{y^2}{2}\;\Hmu, }\\
	 c_{nn} = &\;  \displaystyle{ \frac{\beta^2(-1+y^2)}{4 z^2} \;  [{\hat c}_{VV} \pm \frac{{\hat c}_1}{2}]  } \, ,\\
	 c_{rk} = &\;  \displaystyle{\frac{y \sqrt{1-y^2}}{2 z}\; [{\hat c}_{VV} \pm \frac{{\hat c}_1}{2}]  \:+\: \frac{\beta\sqrt{1-y^2}}{4 z} \;  [{\hat c}_{AA}  \pm \frac{{\hat c}_2}{2} ] \:+\: \frac{y\sqrt{1-y^2} (2-\beta^2)}{4 z}\; \Hmu,} \\
		c_{r} = &\;\displaystyle{ -\frac{y \beta \sqrt{1-y^2}}{2z} \; {\hat c}_{(--)} \:+\: \frac{y\beta\sqrt{1-y^2}}{4z}\;\Hd,}  \\
	c_{k} = & \; \displaystyle{ \frac{1}{2}\beta (1-y^2) \; {\hat c}_{(--)} \:+\: \frac{\beta}{4} (-1+z^2)\; \Hd  }\, .
	\end{eqnarray}
 The terms proportional to $\Hmu$, $\Hd$, and  ${\hat c}_{(--)}$
  apply to all $q\neq t$. The terms proportional to the couplings ${\hat c}_X$ apply only to the
  first quark generation; the upper (lower) sign refers to $u{\bar u}$ ($d{\bar d}$) in the initial state.

\acknowledgments
 We wish to thank C. Deterre, J. Howarth, M. Levy, G. McGoldrick, Y. Peters, M. Rudolph, and  R. Sch\"afer of the ATLAS {\it Top-quark spin analysis group}, and C. Schwanenberger
  for continuing discussions.  The work of  W.B. was supported by BMBF  and that of Z.G. Si  by NSFC and by Natural Science Foundation of
Shandong Province.  D.H. is supported  by Deutsche Forschungsgemeinschaft through Graduiertenkolleg GRK 1675.



\end{document}